\documentclass[prx,twocolumn,showpacs,amsmath,amssymb,superscriptaddress]{revtex4-1}
\usepackage{psfrag,graphicx}
\usepackage{amsfonts,amssymb,amsmath}      
\usepackage{graphicx}
\usepackage{dcolumn}
\usepackage{bm}
\usepackage{float}
\usepackage{hyperref}
\usepackage{accents}
\usepackage{bbding}
\usepackage{color,soul}
\usepackage{epstopdf}
\usepackage{changepage}
\usepackage[normalem]{ulem}

\newcommand{\erf}[1]{Eq.~(\ref{#1})}
\newcommand{\beq}{\begin{equation}}
\newcommand{\eeq}{\end{equation}}
\newcommand{\nn}{\nonumber}

\newcommand{\erfs}[2]{Eqs.~(\ref{#1})--(\ref{#2})}

\newcommand{\dg}{^\dagger}

\newcommand{\smallfrac}[2]{\mbox{$\frac{#1}{#2}$}}

\newcommand{\half}{\smallfrac{1}{2}}
\newcommand{\bra}[1]{\langle{#1}|}
\newcommand{\ket}[1]{|{#1}\rangle}
\newcommand{\ip}[2]{\langle{#1}|{#2}\rangle}

\newcommand{\ito}{It\^o}

\newcommand{\Tr}{\text{Tr}}

\newcommand{\tp}{^{\top}}

\newcommand{\s}[1]{\hat{\sigma}_{#1}}
\renewcommand{\c}{_{\text{C}}}
\newcommand{\ob}{_{\text{o}}}
\newcommand{\un}{_{\text{u}}}

\newcommand{\ex}[1]{\langle{#1}\rangle}
\newcommand{\dd}{{\rm d}}

\newcommand{\ie}{{\em i.e.}}

\newcommand{\past}[1]{\overleftarrow{#1}}
\newcommand{\fut}[1]{\overrightarrow{#1}}
\newcommand{\both}[1]{\overleftrightarrow{#1}}
\newcommand{\fil}{_{\text F}}
\newcommand{\rfil}{_{\text R}}
\newcommand{\sm}{_{\text S}}
\newcommand{\swv}{_{\rm SWV}}
\newcommand{\god}{_{\text T}}
\newcommand{\inv}{^{-1}}
\newcommand{\bx}{{\bf x}}

\newcommand{\by}{{\bf y}}

\newcommand{\est}[1]{\check{#1}}

\newcommand{\cl}{^{\rm cl}}
\renewcommand{\ss}{^{\rm ss}}

\definecolor{nblue}{rgb}{0.06,0.3,0.73}
\definecolor{nblack}{rgb}{0,0,0}
\definecolor{nred}{rgb}{0.9,0.1,0.1}
\definecolor{nmagenta}{rgb}{0.7,0.0,0.3}

\newcommand{\blk}{\color{nblack}}

\definecolor{applegreen}{rgb}{0.55, 0.71, 0.0}
\newcommand{\applegreen}{\color{applegreen}}

\newcommand\stPW{\bgroup\markoverwith{\applegreen{\rule[0.5ex]{2pt}{0.4pt}}}\ULon}
\newcommand{\frf}[1]{Fig.~\ref{#1}}

\begin{document}

\title{Quantum state smoothing cannot be assumed classical \\ even when the filtering and retrofiltering are classical}

\author{Kiarn T. Laverick}
\email{dr.kiarn.laverick@gmail.com}
\affiliation{Centre for Quantum Computation and Communication Technology 
(Australian Research Council), \\ Centre for Quantum Dynamics, Griffith 
University, Yuggera Country, Brisbane, Queensland 4111, Australia}
\author{Prahlad Warszawski}
\affiliation{Centre of Excellence in Engineered Quantum Systems (Australian
Research Council),\\ School of Physics, The University of Sydney, Gadigal Country, Sydney, 
New South Wales 2006, Australia}
\author{Areeya Chantasri}
\affiliation{Centre for Quantum Computation and Communication Technology 
(Australian Research Council), \\ Centre for Quantum Dynamics, Griffith 
University, Yuggera Country, Brisbane, Queensland 4111, Australia}\affiliation{Optical and Quantum Physics Laboratory, Department of Physics,
Faculty of Science, Mahidol University, 
Bangkok 10140, Thailand}
\author{Howard M. Wiseman}
\email{prof.howard.wiseman@gmail.com}
\affiliation{Centre for Quantum Computation and Communication Technology 
(Australian Research Council), \\ Centre for Quantum Dynamics, Griffith 
University, Yuggera Country, Brisbane, Queensland 4111, Australia}

\date{\today}
\begin{abstract}

State smoothing is \blk a technique to estimate a state at a particular time, conditioned on 
information obtained both before (past) and after (future) that time. 
\blk For a classical 
\blk system, the smoothed state
is a normalized product of the {\em filtered \blk 
state} (a state conditioned only on the past measurement information and 
the initial preparation) and the {\em retrofiltered effect} (depending only
on the future measurement information). For the quantum case, \blk whilst \blk there are 
well-established analogues of the filtered state ($\rho\fil$) and 
retrofiltered effect ($\hat E\rfil$), their product does not, in general, 
provide a valid quantum state for smoothing. However, this procedure does 
seem to work when $\rho\fil$ and $\hat E\rfil$ are mutually diagonalizable.
This fact has been used to obtain smoothed quantum states --- more pure 
than the filtered states --- in a number of experiments on continuously 
monitored quantum systems, in cavity QED and atomic systems. In this paper 
we show that there is an implicit assumption underlying this technique: 
that if all the information were known to the observer, the true system 
state would be one of the diagonal basis states. This assumption does not 
necessarily hold, as the missing information is quantum information. It 
could be known to the observer only if it were turned into a classical 
measurement record, but then its nature depends on the choice of 
measurement. We show by a simple model that, depending on that measurement 
choice, the smoothed quantum state can: agree with that from the classical 
method; disagree with it but still be co-diagonal with it; or not even be 
co-diagonal with it. That is, just because filtering and retrofiltering 
appear classical does not mean classical smoothing theory is applicable in 
quantum experiments.

\end{abstract}
\maketitle

\blk\section{Motivation and synopsis}\blk

The true state of a stochastically evolving system will typically be 
unknown, even if the system is continuously monitored through time, because
of information that the observer is missing. In this situation, the true 
state at time $t$ could be estimated by {\em filtering} --- using the 
record $\past{\bf O}_t$ before $t$ --- or by {\em retrofiltering} --- using
the record $\fut{\bf O}_t$ after $t$. However, the best estimate comes from
{\em smoothing}, using the entire record, $\both{\bf O}$. For classical 
systems, these estimates are, most generally, expressed as \blk probability distributions $\wp$ involving the event $\bx =\bx\god$, the unknown true state, and the \blk  smoothed 
distribution is simply related to the others: 
\beq\label{CSS}
\wp\sm(\bx;t) = \wp(\bx;t|\both{\bf O}) \propto 
E\rfil(\bx;t)\wp\fil(\bx;t)\,.
\eeq
Here $\wp\fil(\bx;t) := \wp(\bx;t|\past{\bf O}_t)$ is called the filtered 
state and $E\rfil(\bx;t) \propto \wp(\fut{\bf O}_t|\bx,t)$ is the 
retrofiltered effect. \blk In quantum state estimation, there are well 
established analogues: the filtered state $\rho\fil(t)$ and 
the retrofiltered effect (POVM element) $\hat E\rfil(t)$, where we 
borrowed this terminology in the classical case for consistency. \blk
However, quantum smoothing is not so straight-forward.

If $\rho\fil$ is not pure, that implies that there is information missing 
to the observer. Thus, intuitively, it should be possible to obtain a 
better estimate by smoothing, by using $\both{\bf O}$ instead of just  
$\past{\bf O}_t$. Naively following the classical form (\ref{CSS}) and 
multiplying $\rho\fil(t)$ and $\hat E\rfil(t)$ does not, in general, lead 
to a valid quantum state \cite{GueWis15}. However, there is one case where it does: when 
$\rho\fil(t)$ and $\hat E\rfil(t)$ are co-diagonal in some basis. Then 
multiplying these quantities replicates exactly the classical smoothing 
calculation (\ref{CSS}), with the diagonal elements in these matrices 
acting like classical probabilities for the true state.

The applicability of classical smoothing theory for quantum systems where 
both the filtering and retrofiltering have this classical description seems
quite reasonable. It has been used in a number of experiments on quantum 
systems where the diagonal states are atom-field dressed states 
\cite{Mabuchi09,Mabuchi11}, atomic energy levels~\cite{GamMol14}, and 
photon number states~\cite{RybMol15}. But can it really be justified?  

In this paper we show that classical smoothing gives the best estimate for 
the true quantum state only with an extra assumption: that the missing 
information is such that, if it were known, the state would be in one of 
the diagonal basis states. Whilst  this assumption may seem plausible, it is not entailed by the dynamics. If the system is 
quantum then the missing information is also quantum. For it to be knowable, in principle, it must be turned into a classical measurement record, a second record alongside the observer's record $\both{\bf O}$. Then the 
smoothed quantum state~\cite{GueWis15,LCW19,LCW-QS21, 
LCW-PRA21,Laverick21}, \blk denoted   $\rho\sm$, \blk can be defined as the optimal estimate 
\cite{LGW-PRA21,CGLW21} of the state conditioned on both 
measurement records (\ie, \blk the true state,  denoted  $\rho\god$) by the observer who knows $\both{\bf O}$ but to whom the second record is unknown. Crucially,  the nature of this second measurement record depends on the type of detector which is assumed to create that record from the missing quantum information in the system's environment.  

We consider three different ways to perform the second measurement, to \blk prove various results relevant to our argument.  
In all cases, we use the same simple system, and the primary observer performs the same type of measurement, for which $\rho\fil$ and $\hat E\rfil$ are co-diagonal.  \blk
For the first measurement choice, the \blk true \blk state $\rho\god$ is a pure state in the co-diagonal basis of $\rho\fil$ and $\hat E\rfil$, and $\rho\sm$ reproduces the classical smoothed state, as expected. For the second choice, $\rho\god$ is not in this  diagonal basis, and a different $\rho\sm$ is obtained, albeit one which is still
diagonal in this basis. For the third choice, $\rho\god$ is again not in this  diagonal basis, but this time the smoothed 
state $\rho\sm$ is also not co-diagonal with the classically obtained smoothed state. We further show that, contrary to what one might expect, the classically obtain smoothed state is not even the most optimal, in terms of the expected value of the cost function which defines the smoothed quantum state. 
Our results highlight and delineate the limitations of \blk 
applying classical estimation techniques to quantum systems even when they 
seem adequate.

\blk\subsection{Outline}\blk
In Sec.~\ref{sec-background}, we review classical state smoothing in the classical setting as well as how it has been applied to quantum system. We then discuss some actual experimental examples which have made use of the classical smoothing theory and before introducing the quantum state smoothing theory. 
Following this, in Sec.~\ref{sec-results}, we summarize the \blk three \blk main results of this paper and provide a rough outline of the arguments that lead to them. 

Next, in Sec.~\ref{sec-ex}, we introduce the physical system 
that will serve to prove our results. 
In Sec.~\ref{ssec-ex_fil}, \blk we specify the type of measurement (by Alice, say) on the system outputs that will yield the observed records on which all of the state estimates are conditioned. \blk  Alice's measurements are chosen such that the filtered state and retrofiltered effect are co-diagonal, the situation in which classical smoothing theory would seem appropriate. \blk 
In Sec.~\ref{ssec-photo}, we consider the system outputs that are not accessible to Alice. We specify a measurement (by Bob, say) on these such that the optimal estimate by Alice, using quantum state smoothing~\cite{LGW-PRA21,CGLW21},  coincides with that obtained by classical smoothing. \blk Before moving on, \blk we  compare this smoothed quantum state to the 
filtered state, to attain a deeper understanding of the physical system in 
this regime. \blk We also resolve what may seem a puzzling feature of  
the expected cost functions that define these optimal states, to lay the groundwork for later results on this subject in this paper.\blk  

The subsequent three sections in this paper are dedicated to proving \blk its three main results. \blk 
In Sec.~\ref{ssec-hom}, by changing Bob's 
measurement scheme to a homodyne measurement, we show
that the smoothed quantum state does not \blk necessarily \blk reduce to the classically smoothed
state independent of the \blk choice of \blk secondary measurement 
\blk (Result 1). \blk In Sec.~\ref{ssec-adap}, we again change Bob's 
measurement strategy, \blk this time to \blk an adaptive 
measurement scheme, 
to show that the smoothed quantum state is not \blk even necessarily co-diagonal with the filtered state and \blk retrofiltered effect (Result 2). 
Next, in Sec.~\ref{sec-R3}, we compare which of these three 
cases yield the lowest expected cost function, and hence could be the most 
desirable choice of assumed measurement for Bob, showing that 
for the majority of the time the classical case performs the worst (Result 3). Lastly, 
in Sec.~\ref{sec-concl}, we conclude the paper, and provide some questions for future research.

\section{Classical and Quantum \\ State Smoothing}\label{sec-background}

\subsection{Classical State Smoothing}\label{ssec-css}
For ease of presentation, we will consider classical systems that can be described by a countable number of discrete-valued parameters, collected in a vector $\bx \in \mathbb X$, called the configuration. 
The characteristic of classical systems is that there exists a true or objective configuration $\bx\god(t)$, defining definite values for these parameters. It is only through one's lack 
of knowledge about the initial configuration of the system and the environment with which it interacts 
that this value is obscured. As such, the best description one can give about
the parameters is through the state $\wp(\bx;t)$, a non-negative distribution over possible true configurations normalized such that $\sum_{{\bf x}\in \mathbb X}\wp(\bx;t) = 1$. This state in general has non-zero entropy; to use quantum terminology it is a mixed state, unlike the pure state $\wp\god(\bx;t) = \delta_{\bx,\bx\god(t)}$ corresponding to the true configuration. 

Often, the mixedness of a state will increase over time due to interaction with the environment. However, through 
measurement, which we will take to be a continuous-in-time measurement, one can gain 
information about the hidden true state. 
Note that, since we \blk are only considering discrete classical systems\blk, the dynamics of the configuration are restricted to transitions between the points in $\mathbb X$. We will 
further restrict to Markovian systems; that is, both the measurement results at time $t$ and the state of the system at time $t+\dd t$ are  governed only by the state at the current time, without the need to specify any of the states prior to that. This model of classical systems is 
commonly referred to as a hidden Markov model, as the true state underlying the observed measurement results is hidden. 

Given information about the system, in the form of a measurement record $\bf O$, one can ask:  what is the best estimate $\check\wp(\bx;t)$ of the true state using that information? Here we use a check (rather than the more common hat) to denote an estimate in order to prevent confusion with quantum operators, for which we will use hats. In order to define this optimal estimate
one needs a measure of closeness between the estimated and true states, 
henceforth referred to as the cost function. In this work, we will consider the 
sum-square deviation cost function,
\beq\label{ssd}
{\cal C}[\check\wp,\wp\god] = \sum_{\bx \in {\mathbb X}} 
[\check\wp(\bx;t) - \wp\god(\bx;t)]^2\,.
\eeq
This cost function is the state analogue of the square error cost 
function that is often employed when directly estimating the configuration 
instead of the state.

The above cost function depends on the true state which is, of course, unknown, since it is what one is trying to estimate. Thus one must consider a way to turn the cost function into a value (to be minimized) that is independent of the true state. The simplest way is 
the Bayesian approach to estimation~\cite{Rob07,Sam10,Par09,Ber13}, in which 
one aims to minimize an {\em expected} cost function given the available 
measurement record, defined as
\beq \label{c-risk}
{\cal B}\c[\check\wp] = \mathbb{E}_{\wp\god|{\bf O}\c}
\{{\cal C}[\check\wp,\wp\god]\}\,.
\eeq
Here, ${\mathbb E}_{X|Y}\{Z\}$ denotes an ensemble average of $Z$ over $X$ 
given $Y$, where as a convention $X$ will be omitted when $X=Z$. We have introduced a dummy subscript `C' (standing for conditioning) on ${\bf O}$ to denote what part of the measurement record is available.
For example, one has a filtered conditioning ${\rm C} = {\rm F}$ if only 
the past measurement record, \ie, ${\bf O}\fil = \past{\bf O}_t = \{\by(\tau):\tau\in[t_0,t) \}$, is 
available. Here $\by(\tau)$ is the measurement outcome (detector click, 
photocurrent, etc.) at time $\tau$ and $t_0$ is the initial time.
Similarly, to have a smoothed conditioning ${\rm C} = {\rm S}$, 
we must have access to the past-future measurement record 
${\bf O}\sm = \both{\bf O} = \{\by(\tau):\tau \in [t_0,T) \}$, where $T$ is the final time. 

It is easy to show~\cite{CGLW21} that the optimal estimator for a 
sum-square deviation 
cost function is  
\beq\label{c-opt}
\wp^{\rm opt}\c(\bx) = \mathbb{E}_{\wp\god|{\bf O}\c}\{\wp\god(\bx;t)\}\,.
\eeq
By substituting in the true state 
$\wp\god(\bx;t) = \delta_{\bx,\bx\god(t)}$ we can simplify the computation 
of \erf{c-opt} drastically, obtaining 
$\wp^{\rm opt}\c(\bx;t) = \wp(\bx;t|{\bf O}\c)$.
Thus, the optimal filtered estimate of the true state is defined as 
\beq \label{C-fil}
\wp\fil^{\rm opt}(\bx;t) = \wp(\bx;t|\past{\bf O}_t)\,. 
\eeq
Similarly, the optimal smoothed estimate is defined as 
\beq\label{C-sm}
\wp\sm^{\rm opt}(\bx;t) = \wp(\bx;t|\both{\bf O}) \propto E\rfil(\bx;t)\, 
\wp\fil(\bx;t)\,,
\eeq
where 
\beq
E\rfil(\bx;t) = {\wp}(\fut{\bf O}_t|\bx,t)\,,
\eeq
is also a non-negative function over possible configurations. 
Borrowing terminology from quantum measurement theory, we call $E\rfil(\bx;t)$ an effect. Specifically, because it involves the measurement record over the whole future, we call it a retrofiltered (R) effect. \erf{C-sm} can be derived  
by applying Bayes' theorem and remembering that the system is Markovian. Thus the optimal smoothed state involves 
both the optimal filtering and the {\blk Bayesian} retrofiltering.

\subsection{Quantizing State Estimation}

In open quantum systems a similar idea to classical state estimation 
exists, where one instead aims to give an 
estimated quantum state, $\rho(t)$, 
based on the outcomes of a continuous-in-time measurement. In 
fact, the ideas of both filtering and retrofiltering translate quite 
naturally to quantum systems. Quantum filtering~\cite{Bel87, Bel92}, also known as quantum trajectory theory~\cite{Carmichael93,WisMil10}, is concerned with determining the conditional 
evolution of the quantum state of an open quantum systems where the 
environment is subject to a continuous-in-time measurement. This 
conditional state is called 
the filtered quantum state $\rho\fil(t) := 
\rho_{\past{\bf O}_t}$, the analogue of $\wp\fil(\bx;t)$, as it conditions the state of the system on all 
measurement outcomes up until the estimation time $t$. 
This analogy is actually very close; the dynamical map in quantum trajectory theory, is an obvious quantization of the dynamical map in classical filtering~\cite{WisMil10}.

As for retrofiltering, the likelihood function 
$\tilde\wp(\fut{\bf O}_t|\bx,t)$ is replaced by a quantum effect, \ie, a 
POVM (positive operator-valued measure) element. In quantum measurement 
theory~\cite{BrePet06,Bar09,WisMil10} a POVM is a set of positive hermitian
operators $\{\hat{E}_{o}\}$ whose elements have an expectation value 
equal to the probability of outcome $o$ occurring, \ie, $\Tr[\hat{E}_o \rho] = \wp(o|\rho)$. From this, it is 
easy to see that the quantum analogue of the classical retrofiltered effect 
is the operator $\hat{E}\rfil(t)$ defined such that 
$\wp(\fut{\bf O}_t|\rho,t) = \Tr[\hat{E}\rfil(t)\rho]$. 

With both filtering and retrofiltering having a direct quantum analogue, 
one would be forgiven for assuming that the smoothing technique also has some 
trivial quantization. Based on the classical formula for smoothing in \erf{C-sm}, one 
might define the smoothed quantum state as
\beq\label{swv}
\varrho\swv(t) = \frac{\hat{E}\rfil(t)\circ\rho\fil(t)}{\Tr[\hat{E}\rfil(t)
\circ\rho\fil(t)]}\,,
\eeq
where the Jordan product, defined as $A\circ B = \half(AB + BA)$, is used 
to ensure that the operator is Hermitian, and the denominator is for 
normalization. The subscript `SWV' will be made clear shortly. However,
due to the fact that the filtered state and retrofiltered 
effect do not commute, in general, this construction does not {\blk always}
yield a valid (\ie, positive semi-definite) quantum 
state. As such, one cannot use it as the general definition of the smoothed 
quantum state. In fact, this operator has a close connection to weak-values (see 
Ref.~\cite{CGLW21} for an in-depth discussion), 
and we refer to it, following Refs.~\cite{CGLW21,Ohki22} 
as the {\em smoothed weak-value} (SWV) state.

\subsection{Quantum Experiments Involving \\Smoothing of the State} \label{sec:exp}

\blk Whilst \blk the na\"ive quantization (\ref{swv}) of the smoothed quantum state does 
not generally yield a valid estimate, state smoothing has been applied with 
apparent success in several quantum experiments. All of them were done in 
cavity QED, a platform known for high efficiency continuous measurements 
and superb control of individual quantum systems \cite{MabDoh02}. They all 
succeeded \blk in obtaining a valid quantum state\blk~when applying {\em classical} 
smoothing techniques to the quantum system. Here we summarize these experiments. 
\blk Note that it is not essential for the reader to understand these details in order to grasp the key results 
of this paper.\blk

The first experiment is that reported in 
Refs.~\cite{Mabuchi09,Mabuchi11}. This experiment involves dropping Caesium
atoms through a driven optical cavity
containing a small number ($10$--$100$) of photons. Its aim was to experimentally witness the bistability~\cite{MabWis98} of
the atom-field dressed state; that is, the tendency of the joint state of a single atom and field, in the strong coupling limit, to rapidly relax to two 
very-nearly orthogonal states, switching between them as in a so-called random telegraph signal \cite{BusPar14,KleJur22}. To demonstrate this 
phenomenon, the authors estimated the phase-quadrature of the intracavity field using the measurement record arising from homodyne detection of the 
cavity output beam.  
To obtain the best estimate, the authors used the entire (past-future) record. 
Specifically, inspired by Ref.~\cite{MabWis98}, they applied classical smoothing to a simplified hidden Markov model with three states, each 
corresponding to a different conditional 
expectation value of the phase-quadrature of the intracavity field. The positive and negative values correspond to the two different dressed-states of 
the atom and field, while the state with a zero-value for the phase-quadrature corresponded to the case (not considered in Ref.~\cite{MabWis98}) where 
no atom was present in the cavity. Another simplification they made was not to calculate the full classical smoothed state in 
\erf{C-sm}. Instead, they plot just the most-likely of the three states given the 
past-future record.

In the second experiment we consider, Ref.~\cite{GamMol14}, a single Caesium atom was trapped within an optical high-finesse cavity, in the strong 
coupling regime. The aim of the experiment was to estimate the occupation probabilities of two possible energy eigenstates of the Caesium atom as well 
as estimating the transition rates between these states. Similarly to the previous experiment, they apply classical state smoothing, where the hidden 
Markov model assumed in this case comprises the two possible energy eigenstates and a third state introduced to help model the additional energy 
levels of the atom. To obtain this estimate, they probed the cavity with an on-resonant (with the empty cavity) weak laser, contributing a small number 
(less than 1) of photons. The output beam was measured by a single-photon detector, with the sequence of detection times forming the measurement record. 
The authors processed this measurement record using classical state smoothing, \erf{C-sm}, to obtain estimates of the occupation probabilities of the 
energy eigenstates. 

The final experiment
Ref.~\cite{RybMol15}, also involves a high-finesse cavity and atoms, but the relevant cavity mode is at microwave frequency, and the focus of attention 
is on the dynamics of its state. The atoms, which are Rubidium atoms in circular Rydberg states, are used just to probe the number of photons in the 
cavity. Specifically, 
the Rydberg atom is prepared in a 
superposition of circular Rydberg eigenstates $\ket{+} = (\ket{50} + \ket{51})/\sqrt{2}$,  
where $\ket{50(51)}$ is a circular state with principle quantum number 50(51),
through an 
interaction with a separate low-finesse microwave cavity. These atoms then 
interact with the intracavity field, causing a photon-number dependent 
relative phase shift $\phi$ between $\ket{50}$ and $\ket{51}$. As the atom 
exits the cavity, it interacts with a second low-finesse cavity which 
rotates the atomic superposition with some particular relative phase $\phi'$ to $\ket{50}$ (and the orthogonal state to $\ket{51}$) after which the 
atomic state is measured projectively. By continuously (at least to a good 
approximation) probing the system with Rydberg atoms, one forms the 
measurement record from the outcomes of the projective measurement. 
However, due to the periodicity of the relative phase shift, this 
measurement can only detect the photon number modulo $k$, where for this
experiment $k = 8$. The authors used    
this measurement record to estimate the photon number probabilities via 
the classical smoothing theory in \erf{C-sm}, and showed an improvement (higher purity) relative to the estimate obtained using classical 
filtering theory, \erf{C-fil}. Their hidden Markov model is equivalent to assuming that the intracavity field is in some Fock state $\ket{n}$.

\subsection{Why does classical smoothing seem to work in these experiments?} \label{sec:whyexp}

In all of the above experiments, the authors apply 
classical smoothing theory to systems that are, undoubtedly, quantum, while still obtaining sensible results. 
Although none of these papers 
explicitly writes down their estimate as a quantum state, there 
is a trivial relationship between the smoothed estimates of the 
probabilities, $\wp\sm(x;t)$ and the corresponding 
estimate of the state: $\est\rho(t) = \sum_{x} \wp\sm(x;t) 
\ket{\psi_x}\bra{\psi_x}$. Here the set of pure states $\{\ket{\psi_x}\}$ 
are those assumed in their respective hidden Markov models: dressed atom-field states~\cite{Mabuchi09,Mabuchi11}, 
atomic energy states~\cite{GamMol14}, and Fock states~\cite{RybMol15}). We now examine in detail what makes classical smoothing applicable in these cases.  

In all these experiments the quantum states corresponding to the $d$ discrete states in the hidden Markov model are {\em orthogonal}, or very-nearly so. 
This means that, in the orthonormal basis 
$\{\ket{\psi_x}:\langle\psi_x|\psi_{x'}\rangle = \delta_{x,x'} \}_{x = 1}^d$, the `true' quantum state of the system at any given
time is \beq
\rho\god(t) = \sum_{x = 1}^d \wp\god(x;t)\ket{\psi_x}\bra{\psi_x}\,,
\eeq
with $\wp\god(x;t) = \delta_{x,x\god(t)}$. The important point is that the 
true state will always be diagonal in this basis, which means that so will 
the filtered state and retrofiltered effect (See App.~\ref{app-qest} for the proofs). That is, 
\begin{subequations}\label{eq-mutdiag}
\begin{align}
\rho\fil(t) &= \sum_{x = 1}^d \wp(x;t|\past{\bf O}_t)
\ket{\psi_x}\bra{\psi_x}\,,\\ 
\hat{E}\rfil(t) &= \sum_{x = 1}^d \wp(\fut{\bf O}_t|x,t)\ket{\psi_x}\bra{\psi_x}\,,
\end{align}
\end{subequations}
which trivially commute, $[\rho\fil(t),\hat{E}\rfil(t)] = 0$. Since the 
problem of the smoothed weak-valued state becoming indefinite only occurs 
when $[\rho\fil(t),\hat{E}\rfil(t)] \neq 0$, the orthogonality assumption 
removes any possibility of $\varrho\swv(t)$ becoming physically invalid at 
any point in the evolution. In this case, the SWV state becomes
\beq\label{eq-clqtsmooth}
\rho\sm\cl(t) = \sum_{x = 1}^d \wp\sm(x;t)\ket{\psi_x}\bra{\psi_x}\,,
\eeq
where $\wp\sm(x;t)$ is defined in Eq.~\eqref{C-sm}. \blk From this 
point forward, we will refer to this as the `classical regime', hence the 
superscript `cl'.

While the assumption Eqs.~\eqref{eq-mutdiag} and \eqref{eq-clqtsmooth} 
facilitates the usage of the SWV state as a smoothed estimate of the 
quantum state, it has also effectively removed the `quantumness' from the 
system as it assumes that all the information that was missed by the observer 
was classical in nature. 
Moreover, there is an obvious tension in this semi-classical treatment.  
On the one hand, a portion of the information in the
environment is treated as quantum information, \ie, that portion captured 
by the observer, whose choice of measurement affects how 
it is converted into classical information. On the 
other hand the remaining portion left in the environment is 
treated as purely classical, 
revealing which of the orthogonal basis states the system is in. To 
obtain a more consistent treatment of quantum systems, all the information 
in the environment should be treated 
on the same footing, being subject to different measurement choices (``unravellings''). 
But to deal with this generalization requires the quantum state smoothing theory of
Ref.~\cite{GueWis15}.

\subsection{Quantum State Smoothing}

Similar to the classical state smoothing in Section~\ref{ssec-css}, 
the quantum state smoothing theory \cite{GueWis15} assumes a hidden 
Markov model with possible true (unknown) states, 
$\rho\god(t)$, consisting of only valid quantum states of the system. 
Importantly, these possible true states are not restricted to only 
orthogonal basis states as in the classical-like quantum smoothing. The 
goal is to obtain an estimated state $\est\rho(t)$ that is closest to 
possible true states. As a quantum analogue of the sum-square deviation 
considered for classical systems, we consider a trace-square deviation 
expected cost function
\beq
{\cal B}_{{\bf O}\c}^{\rm TrSD}[\est\rho] = 
{\mathbb E}_{\rho\god|{\bf O}\c}\left\{\Tr\left[(\est\rho(t) - 
\rho\god(t))^2\right]\right\}\,,
\eeq
where the average is taken over the set of possible true states 
${\mathbb T}_t$. The estimator that 
minimizes this expected cost is~\cite{LGW-PRA21}
\beq
\rho\c = {\mathbb E}_{\rho\god|{\bf O}\c}\{\rho\god(t)\}\,,
\eeq
attaining the minimum value
\beq\label{simp-risk}
{\cal B}_{{\bf O}\c}^{\rm TrSD}[\rho\c] = 
{\mathbb E}_{\rho\god|{\bf O}\c}\{P[\rho\god(t)]\} - P[\rho\c(t)]\,,
\eeq
where $P[\rho] = \Tr[\rho^2]$ is the purity of a state $\rho$. If
the conditioning `C' refers to the past observed record $\past{\bf O}_t$, 
the optimal estimate is the filtered state
$\rho\fil(t) := {\mathbb E}_{|\past{\bf O}_t}\{\rho\god(t)\}$. 
This is identical to the filtered state defined in standard quantum trajectory theory. 
When the conditioning is on both past and future records, 
$\both{\bf O}$, the optimal estimate of the true state is 
\beq\label{QSS}
\rho\sm(t) = {\mathbb E}_{|\both{\bf O}}\{\rho\god(t)\}\,.
\eeq
This is the definition of the smoothed quantum state, which, unlike the SWV 
state, is guaranteed to be a valid quantum state as it is a convex 
combination of valid quantum states. 

It should be clear that, in general, it is the case that 
$\rho\sm(t) \ne \rho\sm\cl$, due to the non-orthogonality of the set of 
true states ${\mathbb T}_t$. However, in the event that the possible true 
states are pure and mutually orthogonal, \ie, when 
${\mathbb T}_t = \{\ket{\psi_x}\bra{\psi_x}: \langle\psi_x|
\psi_{x'}\rangle = \delta_{xx'}\}_{x=1}^d$, we
have a `quantum-classical equivalence' 
$\rho\sm (t) = \rho\sm\cl(t)$~\cite{LCW-QS21}. Note, 
this is only a sufficient condition for a quantum-classical equivalence. 
However, at this point we have yet to specify how one can obtain the set of 
possible true states. Unlike in the classical state estimation where possible true states can be 
directly associated with true configurations $\bx\god(t)$ of the system, in
the quantum state estimation, the set of possible true states 
${\mathbb T}_t$ is more subtle to define. 

It is almost always the case 
that the observer, henceforth referred to as Alice, does not have complete access to the environment into which the 
system is leaking information. (If she did then the filtered state would, typically, 
be a pure state and no decrease in uncertainty from smoothing would be possible while 
maintaining a valid quantum state.) Thus we consider a partition of 
the system's environment or baths into two parts. This applies even if, by conventional accounting, 
there is only one bath, if Alice's detection has some non-unit efficiency $\eta$. 
From this fraction of the bath, Alice's choice of measurement, ${\cal M}_A$, yields 
the measurement record $\bf O$, the `observed' record. As for 
the remaining fraction, $1-\eta$~\cite{Comment1}, this quantum information 
is \emph{unobserved} by Alice. However, as it propagates away from the system, 
the information will encounter more complex environments which induce effectively irreversible 
decoherence. This defines a preferred basis \cite{Zurek03,Strasberg23}
which can be regarded as a choice of measurement, ${\cal M}_B$ of the information, yielding a 
second measurement record ${\bf U}$ that is unobserved by Alice, called the `unobserved' record.
We anthropomorphize this process by calling this Bob's measurement and record. 

With both of these measurement records now defined, it is easy to see that 
the true state of the system is given by 
$\rho\god(t) = \rho_{\past{\bf O}_t \past{\bf U}_t}$, as together the
observed and unobserved measurement records contain the maximum amount of 
information about the system. This true state can be computed
with standard quantum trajectory theory with measurements on multiple baths
\cite{Carmichael93, WisMil10}. The set of possible true states is then
determined by the possible measurement records that could have occurred for
Alice and Bob, given their respective measurement choice, \ie, 
$\mathbb{T}_t = \{\rho_{\past{\bf O}_t \past{\bf U}_t}|{\cal M}_A, 
{\cal M}_B\}_{\past{\bf O}_t \past{\bf U}_t}$. It is with this set that the
smoothed quantum state has been computed with in previous works 
\cite{GueWis15,LCW-QS21,CGLW21,LGW-PRA21}. 

One important thing to notice is that this definition for the set of true 
states explicitly depends on both Alice's and Bob's choice of measurement, \blk as indicated by the notation 
$\mathbb{T}_t = \{\rho_{\past{\bf O}_t \past{\bf U}_t}|{\cal M}_A, 
{\cal M}_B\}_{\past{\bf O}_t \past{\bf U}_t}$. \blk 
\blk This dependence is expected, as we are dealing 
with quantum information, and it 
raises \blk the question of how Bob's choice is determined. As stated above, we may consider Bob to be an anthropomorphic representation 
of the environment, rather than a real agent. In that case it is not possible to `ask' Bob what measurement he performed. 
Rather, one would have to infer the best model for how, ultimately, the quantum 
information from the system is turned into classical information by decoherence 
in the environment. 
However, for experimentally testing the theory, it would be necessary to have a `real' Bob --- an agent who can potentially make different measurement 
choices ${\cal M}_B$, and reveal them to Alice (even while keeping the results from her.) This is the situation we will consider in the remainder of the 
paper.  We stress again that \blk whilst  the \blk choice of unravelling for Bob has no impact on the filtered 
quantum state, this choice \blk can cause drastically different dynamics for the smoothed state \cite{GueWis15,CGW19,LCW-PRA21}. 

\section{Classical-like filtering and retrofiltering does not guarantee classical-like smoothing}\label{sec-results}

With the necessary background covered, we can move onto the main 
question this paper addresses. Say that for a given observed record $\both{\bf O}$, 
Alice's filtered state and retrofiltered effect commute at all times, as in the experiments 
discussed above (see Sec.~\ref{sec:whyexp}). Given only this, is 
it correct for Alice to use the classically smoothed estimate as her {\em optimal} 
smoothed estimate? If this proves true it would open an entire regime 
where quantum state smoothing can be implemented without the need to  
know the measurement ${\cal M}_B$ which Bob performed.
In particular, it might justify the approach already applied in the experiments detailed 
in Sec.~\ref{sec:exp}. 
However,
in the following sections we answer this question, and related questions, definitively in the negative. 
We address this in three stages, from which we obtain the 
following three results:\\

\noindent
\underline{\bf Result 1:} {\em The commutativity of the filtered state and 
retrofiltered effect is \underline{not} sufficient for the smoothed quantum
state to reduce to its classical counterpart at time $t$. That is, 
$[\rho\fil(t),\hat{E}\rfil(t)] = 0 \nRightarrow \rho\sm(t) = \rho\sm\cl(t)$.}\\

\noindent
\underline{\bf Result 2:} {\em The commutativity of the filtered state and 
the retrofiltered effect is \underline{not} sufficient  for the smoothed 
state to mutually commute with both the filtered state and retrofiltered 
effect.}\\

\noindent
\underline{\bf Result 3:} {\em The classical smoothed quantum state does 
not even always have the smallest expected cost function when compared to 
the smoothed state resulting from other unravellings for the environment.}\\

We prove all of these results 
using a single, very simple example open quantum system: a 
single two-level system (qubit) coupled to bosonic baths, the details of 
which are given in Sec.~\ref{sec-ex}. We then fix Alice's 
measurement choice (Sec.~\ref{ssec-ex_fil}) so that the filtered state and retrofiltered effect 
commute over a known time interval $[t_1,t_2)$. This is to ensure that we 
are always operating within the regime of interest for the question. As 
for Bob's measurement, we investigate three potential choices for this. 
The first choice (Sec.~\ref{ssec-photo}) gives the classical regime, where 
his measurement is such that, over the interval $(t_1, t_2)$, $\rho\god(t) 
\in\{\ket{e}\bra{e}, \ket{g}\bra{g}\}$, with $\ip{e}{g} = 0$ causing $\rho\sm(t)$ to equal $\rho\sm\cl(t)$. This case 
is considered first to provide a comparison for the following regimes. 
In the second (Sec.~\ref{ssec-hom}) and third (Sec.~\ref{ssec-adap}) 
cases, Bob's measurement is chosen so that $\rho\god(t)$ is not restricted 
to an orthogonal set 
over the interval $(t_1,t_2)$. 
The second case considers a homodyne measurement of the output and enables us to 
prove result 1. The third case considers an adaptive interferometric measurement strategy 
on the output bosonic field, and enables us to prove result 2. 
As for the final result, 
in Sec.~\ref{sec-R3}, we investigate the expected cost functions of the 
smoothed quantum state in all three cases and show that, for the majority 
of the time, the classical case yields a larger (and hence worse) estimate 
when compared to the other two cases. 

\section{Model: single qubit coupled to emission and absorption channels}
\label{sec-ex}
In this section we introduce the example system that will be used throughout 
the remainder of the paper to prove our results. Also in this section we 
introduce the measurement which Alice makes on her portion of the environment, 
chosen so that the resulting 
filtered state and retrofiltered effect are mutually diagonal over the 
given time frame. This is to ensure that we are always operating within the 
regime where the filtered state and retrofiltered effect are describable by 
a classical discrete hidden Markov model. Finally, in this section, we introduce 
the first choice for Bob's measurement, the one that explicitly realises this 
hidden Markov model and so makes $\rho\sm(t) = \rho\sm\cl(t)$. The two other, 
more complicated, measurement choices for Bob will be explained in the subsequent sections.

The physical system we consider is a qubit that is coupled to three 
decoherence channels: two emission channels and one absorption channel. The
dynamics of the quantum state, under no observation, is specified by a vector of Lindblad operators,
\beq
\hat{\bf c} = (\sqrt{\delta}\s{-},\sqrt{\gamma}\s{-},\sqrt{\epsilon}\s{+}),
\label{Lindblads}
\eeq
This defines the Lindblad master equation 
\beq
\dot\rho = {\cal D}[\hat{\bf c}]\rho \equiv \sum_{\ell} \hat{c}_\ell\rho\hat{c}_\ell\dg - 
\{\hat{c}_\ell\dg\hat{c}_\ell,\rho\}/2,
\label{DecoherenceME}
\eeq
as we are working in a frame where the system Hamiltonian disappears. 
In \erf{Lindblads} we are using the Pauli 
ladder operators $\s{\pm} \equiv (\s{x} \pm i\s{y})/2$, where  
$\s{x,y,z}$ are the usual Pauli operators. We denote the eigenstates of 
$\s{z}$ by $\s{z}\ket{e} = \ket{e}$, $\s{z}\ket{g} = -\ket{g}$. 
The master equation (\ref{DecoherenceME}) is extremely simple in this basis, 
looking like a classical bit stochastically transitioning 
between $\ket{e}$ and $\ket{g}$, which is described by the classical master equation 
for the ground state probability, $\wp(g;t)  = \bra{g}\rho(t)\ket{g}$, 
\beq
\dot{\wp}(g;t) = (\delta + \gamma)\wp(e;t) - \epsilon\wp(g;t)\,,
\eeq
and the probability of being in the excited state being $\wp(e;t) = 1 - \wp(g;t)$.

\subsection{Codiagonal Filtering and Retrofiltering: \\ Alice uses photon detection}
\label{ssec-ex_fil}

Throughout, we consider the case where Alice perfectly monitors the first channel, 
corresponding to emission at rate $\delta$. Since there is a second emission channel, 
with rate $\gamma$, we could imagine a single emission channel which Alice monitors 
with efficiency $\delta/(\delta+\gamma)$. To ensure Alice's filtered state
and retrofiltered effect commute, over some time interval $(t_1,t_2)$, 
it is sufficient to say that Alice uses photon detection to monitor her channel. 
Such a measurement, as 
will be shown shortly, causes both the filtered state and retrofiltered 
effect to share the $\s{z}$-basis as their diagonal basis between any two 
detection events (jumps).  

This measurement results in the following 
stochastic master equation for the filtered state~\cite{WisMil10}, 
\beq\label{rhofil}
\dd\rho\fil = {\cal G}[\hat{c}\ob]\rho\fil\dd N\ob + 
{\cal D}[\hat{\bf c}\un]\rho\fil\dd t - \frac{1}{2}
{\cal H}[\hat{c}\ob\dg\hat{c}\ob]\rho\fil\dd t\,,
\eeq
with $\hat{c}\ob = \sqrt{\delta}\s{-}$, $\hat{\bf c}\un = 
(\sqrt{\gamma}\s{-},\sqrt{\epsilon}\s{+})$. The superoperators 
in \erf{rhofil} are ${\cal G}[\hat{a}]\rho = 
\hat{a}\rho\hat{a}\dg/\Tr[\hat{a}\rho\hat{a}\dg] - \rho$ and 
${\cal H}[\hat{\bf a}]\rho = \sum_k \hat{a}_k\rho + \rho\hat{a}_k\dg - 
\Tr[\hat{a}_k\rho + \rho\hat{a}_k\dg]\rho$, 
and the stochastic increment characterising the jump, $\dd N_{\rm o}$, 
satisfies the following properties
\beq
\mathbb{E}[\dd N_{\rm o}] = \Tr[\hat{c}_{\rm o}\rho\hat{c}_{\rm o}\dg]
\dd t\,, \qquad 
\dd N_{\rm o}^2 = \dd N_{\rm o}\,.
\eeq
Here, we have introduced the subscripts `${\rm o}$' and `${\rm u}$' to 
distinguish the channels observed and unobserved, respectively, by Alice. 

To see that this measurement for Alice does 
result in a filtered state diagonal in the $\s{z}$-basis between any two 
consecutive jumps, we can compute \erf{rhofil} under a ground state initial
condition given by the first jump at time $t_1$, \ie, 
$\rho\fil(t_1^+) = \ket{g}\bra{g}$ \blk with $t_1^+ = t_1 + \dd t$\blk, and a no-jump evolution 
($\dd N\ob(t) = 0$). This results in 
$\rho\fil(t) = \wp\fil(e;t)\ket{e}\bra{e} + \wp\fil(g;t)\ket{g}\bra{g}$ 
with $\wp\fil(g;t)$ satisfying the differential equation
\beq\label{fil_coeff}
\dot{\wp}\fil(g;t) = \gamma\wp\fil(e;t) - \epsilon\wp\fil(g;t) + 
\delta\wp\fil(e;t)\wp\fil(g;t)\,,
\eeq
and $\wp\fil(e;t) = 1 - \wp\fil(g;t)$

As for the retrofiltered effect, the stochastic differential equation 
governing its evolution is~\cite{GueWis15, GueWis20}
\beq\label{Erfil}
\begin{split}
-\dd\hat{E}\rfil = \tilde{\cal G}\left[\hat{c}\ob\dg/\sqrt{\zeta}\right]
\hat{E}\rfil\dd &N\ob + {\cal D}\dg[\hat{\bf c}\un]\hat{E}\rfil\dd t \\ 
&\,\,\,\,\,- \frac{1}{2}\tilde{\cal H}[\hat{c}\ob\dg\hat{c}\ob -\zeta]
\hat{E}\rfil\dd t\,,
\end{split}
\eeq
which is evolved backwards in time from the final condition $\hat{E}\rfil(T) \propto \hat{1}$, 
representing a final uninformative state. 
In \erf{Erfil}, $\tilde{\cal G}[\hat{a}]\rho = \hat{a}\rho\hat{a}\dg - \rho$ and 
$\tilde{\cal H}[\hat{\bf a}]\rho = \sum_k \hat{a}_k \rho + \rho 
\hat{a}_k\dg$ are the linear versions of the nonlinear superoperators 
introduced above, while $\zeta$ is an arbitrary positive constant which affects only 
the norm of $\hat{E}\rfil$. It is related to a so-called ostensible probability distribution 
for jumps; see Ref.~\cite{GueWis20}. Thus, strictly, $\hat{E}\rfil$ is only proportional to the effect, but all the expressions for the smoothed state are normalized so that the norm of $\hat{E}\rfil$ for a fixed $\fut{\bf O}$ does not matter.

Now we show that the retrofiltered effect is also diagonal 
in the $\s{z}$-basis between any two consecutive observed jumps. We have the final 
condition at the time just prior the second jump $t_2$ that 
$\hat{E}\rfil(t_2^-) \propto \ket{e}\bra{e}$, \blk where $t_2^- = t_2 - \dd t$. 
To see this, one can use the quantum map form for the evolution of the retrofiltered effect 
(See App.~\ref{app-qest} for an explanation), 
\beq\label{Eff_map}
\begin{split}
\hat{E}\rfil(t_2^-) = \hat{M}\ob\dg(\by\ob;t_2) \sum_{\by\un} &\hat{M}\un\dg(\by\un;t_2) \hat{E}\rfil(t_2)\times\\
&\hat{M}\ob(\by\un;t_2)\hat{M}\ob(\by\ob;t_2)\,.
\end{split}
\eeq
Since, at $t_2$, a photon is detected in the $\delta$-channel, the 
map corresponding to the observed measurement takes the form $\hat{M}\ob(\by\ob;t_2) \propto \hat{c}\ob \propto \ket{g}\bra{e}$. 
Computing \erf{Eff_map} with this map yields the final condition for the effect\blk. 
Computing the evolution of the 
retrofiltered effect given this final condition at $t_2^-$ and under a 
no-jump evolution ($\dd N\ob = 0$), we obtain the solution 
$\hat{E}\rfil(t) = E\rfil(e;t)\ket{e}\bra{e} + E\rfil(g;t)\ket{g}\bra{g}$, 
where the coefficients satisfy
\beq\label{Cl-Eff}
\begin{split}
\dot{E}\rfil(g;t) &= \epsilon\big(E\rfil(g;t) - E\rfil(e;t)\big) + 
\zeta E\rfil(g;t)\,,\\
\dot{E}\rfil(e;t) &= \gamma\big(E\rfil(e;t) - E\rfil(g;t)\big) + 
(\delta + \zeta) E\rfil(e;t)\,.
\end{split}
\eeq
Thus by Alice choosing photon detection, from her perspective, the 
quantum system can be completely described by a classical hidden 
Markov model between any two detection events. In fact, 
this means that the entire evolution has this nature apart from 
potentially the evolution prior to the first jump, if the initial state 
is not diagonal in the $\s{z}$ basis.

\begin{figure}[t!]
\includegraphics[scale=0.26]{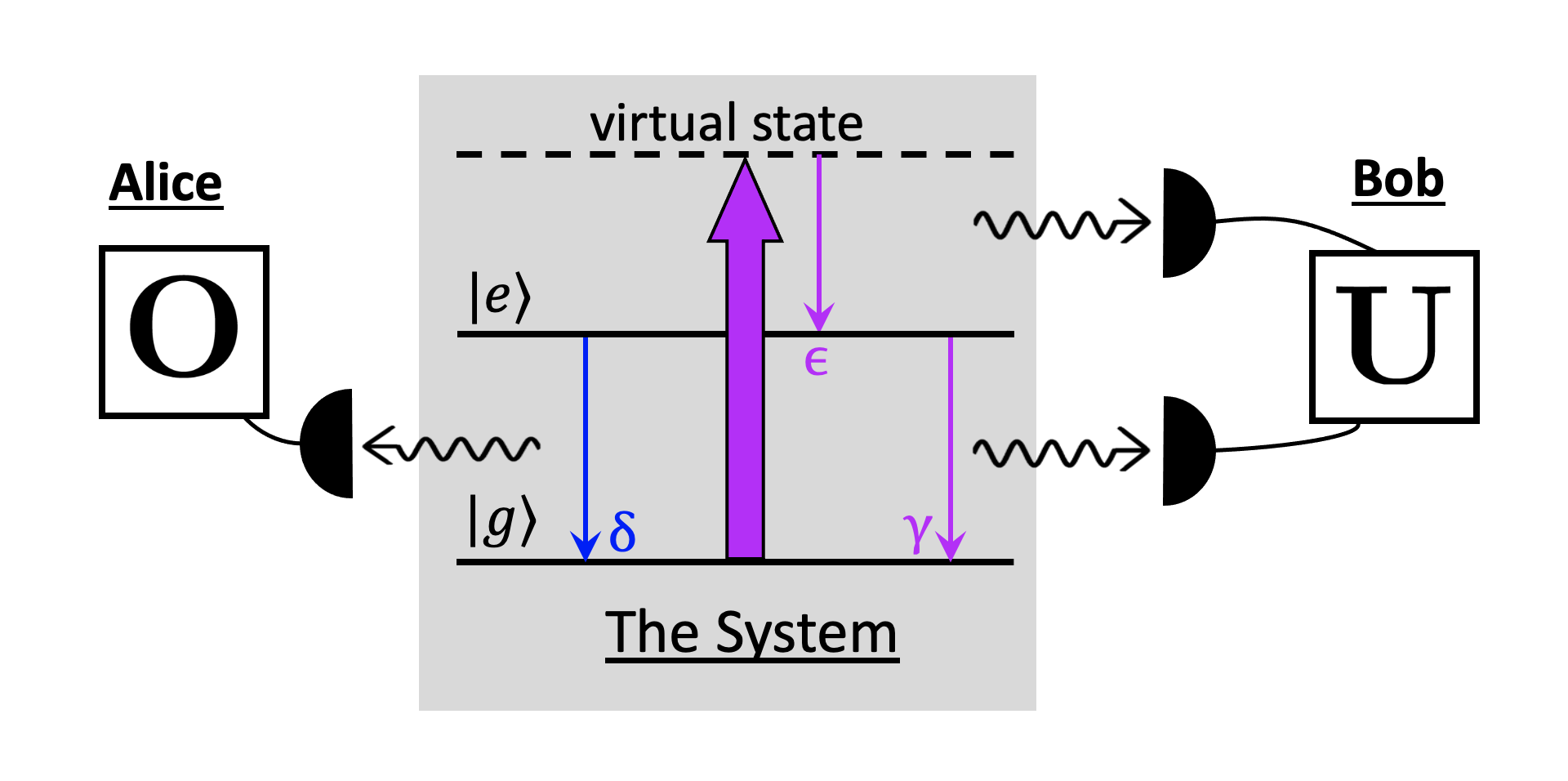}
\caption{A qubit undergoing three separate dynamical processes. The first 
two processes are photon emission processes, one with rate $\delta$ that is
monitored by Alice (blue arrow) using photon detection and one with rate 
$\gamma$ monitored by Bob, again using photon detection. The final process 
is an absorption process with rate $\epsilon$ (dashed arrow), modelled by a
continuous Raman driving (black arrow) to a virtual state that immediately 
decays to the excited state via photon emission, that is observed by Bob 
via photon detection. }
\label{Fig-Photo}
\end{figure}

As Bob monitors all the channels that Alice's 
measurement missed, this leaves him to perfectly monitor the 
$\gamma$- and $\epsilon$-channels. One might be wondering how
it is possible for Bob to measure an absorptive channel. Up
until now we have not provided details about how the absorptive channel 
could have arisen physically. We consider that the qubit is under a 
continuous driving of a Raman transition~\cite{Kuhn02,Kli04,Kel04,Sun18} to a 
virtual state which immediately decays to the excited state by emitting a 
(detectable) photon, see \frf{Fig-Photo}. Such a scheme can be made 
equivalent to an absorptive channel with rate $\epsilon$ while emitting a 
photon of a different frequency to those in the emission channels so that 
it can be \blk monitored, similarly to the emissions channel, by detecting this photon. \blk  
With this technical point taken care of 
we next turn to Bob's first choice of measurement.

\subsection{Classical Smoothed State: Bob\\ uses photon detection}\label{ssec-photo}

In this section we want Bob's measurement to be such that 
${\mathbb T}_t = \{\ket{g}\bra{g},\ket{e}\bra{e}\}$ for all $t\in(t_1,t_2)$, 
with $\rho\god(t_1^+) = \ket{g}\bra{g}$. Just as for Alice, above, 
this is easy to achieve by having Bob monitor 
the remaining two channels perfectly using photon detection. For this 
measurement scheme, the stochastic master equation for the true state 
is~\cite{WisMil10},
\beq\label{SME-jumps}
\begin{split}
\dd\rho\god =&\, {\cal G}[\hat{c}\ob]\rho\god\dd N\ob - 
\frac{1}{2}{\cal H}[\hat{c}\ob\dg\hat{c}\ob]\rho\god\dd t \,+ \\&
{\cal G}[\hat{c}_{{\rm u},\gamma}]\rho\god\dd N_{{\rm u},\gamma} - 
\frac{1}{2}{\cal H}[\hat{c}_{{\rm u},\gamma}
\dg\hat{c}_{{\rm u},\gamma}]\rho\god\dd t \,+ \\
&{\cal G}[\hat{c}_{{\rm u},\epsilon}]\rho\god\dd N_{{\rm u},\epsilon} - 
\frac{1}{2}{\cal H}[\hat{c}_{{\rm u},\epsilon}\dg
\hat{c}_{{\rm u},\epsilon}]\rho\god\dd t\,,
\end{split}
\eeq
where $\dd N_{{\rm u},k}$ are the stochastic increments describing a 
detection of a photon in the corresponding channel. 

To see that this measurement choice of Bob's does indeed result in 
the true state being in either the ground or excited state between 
any two consecutive observed jumps we can look at each term in 
\erf{SME-jumps} individually. Firstly, as we are considering the 
evolution between two consecutive observed jumps, we know that initially 
the true state will be in the ground state and that 
$\dd N\ob(t) = 0$ until the following jump, removing the first term in 
\erf{SME-jumps}. Leaving the terms of order $\dd t$ for last, the remaining
unobserved jump terms project the true state into either the ground or
excited state if a photon is detected in either the $\gamma$- or 
$\epsilon$-channel, respectively. Finally, looking at the terms of order 
$\dd t$, when computing these with the system in the ground state, 
they are equal to zero. This means that the true state will remain in the 
ground state until a photon is detected in the $\epsilon$-channel and 
projects it into the excited state. Computing the $\dd t$ terms also gives 
zero when the true state is in the excited state and it remains unchanged 
until it is projected into the ground state via a photon emission in the 
$\gamma$-channel. Thus, under this monitoring by Bob, the true state 
between any two observed jumps will only be in either the ground state 
$\ket{g}\bra{g}$ or the excited state $\ket{e}\bra{e}$, as claimed.

With ${\mathbb T}_t$ obtained, all that remains is to compute the smoothed 
quantum state. For this case it is a fairly simple task. Beginning with 
\erf{QSS}, we have 
\beq\label{Ph-QSS}
\begin{split}
\rho\sm(t) &= \sum_{m\in\{g,e\}} \wp(m;t|\both{\bf O})\ket{m}\bra{m}\\
& \propto \sum_{m\in\{g,e\}}{\wp}(\fut{\bf O}_t|m,t) 
\wp(m;t|\past{\bf O}_t)\ket{m}\bra{m}\\ 
&= \sum_{m\in\{g,e\}}\Tr\left[\hat{E}\rfil(t)\ket{m}\bra{m}\right] 
\wp(m;t|\past{\bf O}_t)\ket{m}\bra{m}\\
&= E\rfil(g;t) \wp\fil(g;t)\ket{g}\bra{g} + E\rfil(g;t) 
\wp\fil(g;t)\ket{e}\bra{e}\,,\\
\end{split}
\eeq
where we have used Bayes' theorem in the second line and 
in the final line we have recognized that $\wp(m;t|\past{\bf O}_t)$ 
is exactly the coefficient for the filtered state in \erf{fil_coeff}. After
normalization this gives $\rho\sm(t) = \wp\sm(e;t)\ket{e}\bra{e} + 
\wp\sm(g;t)\ket{g}\bra{g} = \rho\sm\cl(t)$, as expected. 

For a simpler analysis, we henceforth consider the case where 
$\delta\to 0^+$. In this limit, Alice very rarely observes a jump 
causing $(t_1,t_2)$ to, typically, be large enough for both the 
filtered state and retrofiltered effect to reach their respective 
stationary solutions between any two consecutive jumps. 
As a result, the filtered state and smoothed 
quantum state only differ for a time on the timescale that the system 
equilibriates, in this case $1/(\gamma + \epsilon)$, prior to the final 
jump. To see why this is the case, we need to look at the retrofiltered 
effect. From \erf{Cl-Eff}, the steady-state solution is
$E\ss\rfil(e) = E\ss\rfil(g) = \lambda(t)$, where $\lambda(t)$ is half the norm of  
$\hat E\rfil$. (Recall that the norm does not affect any calculated quantities, and so can be 
time-dependent even in steady state.) This means that whenever the 
retrofiltered effect is in steady state, the smoothed quantum state in \erf{Ph-QSS}, when 
normalized at time $t$, will be equal to the filtered state. Thus, in the limit  
$\delta\to 0^+$, the retrofiltered effect will be in steady state over the 
entire evolution until a time of order $1/(\gamma + \epsilon)$. Thus we only need 
to consider the evolution of the state on this time scale prior to an observed jump. 

\begin{figure*}[t]
\begin{minipage}{0.3333\textwidth}
\includegraphics[scale = 0.25]{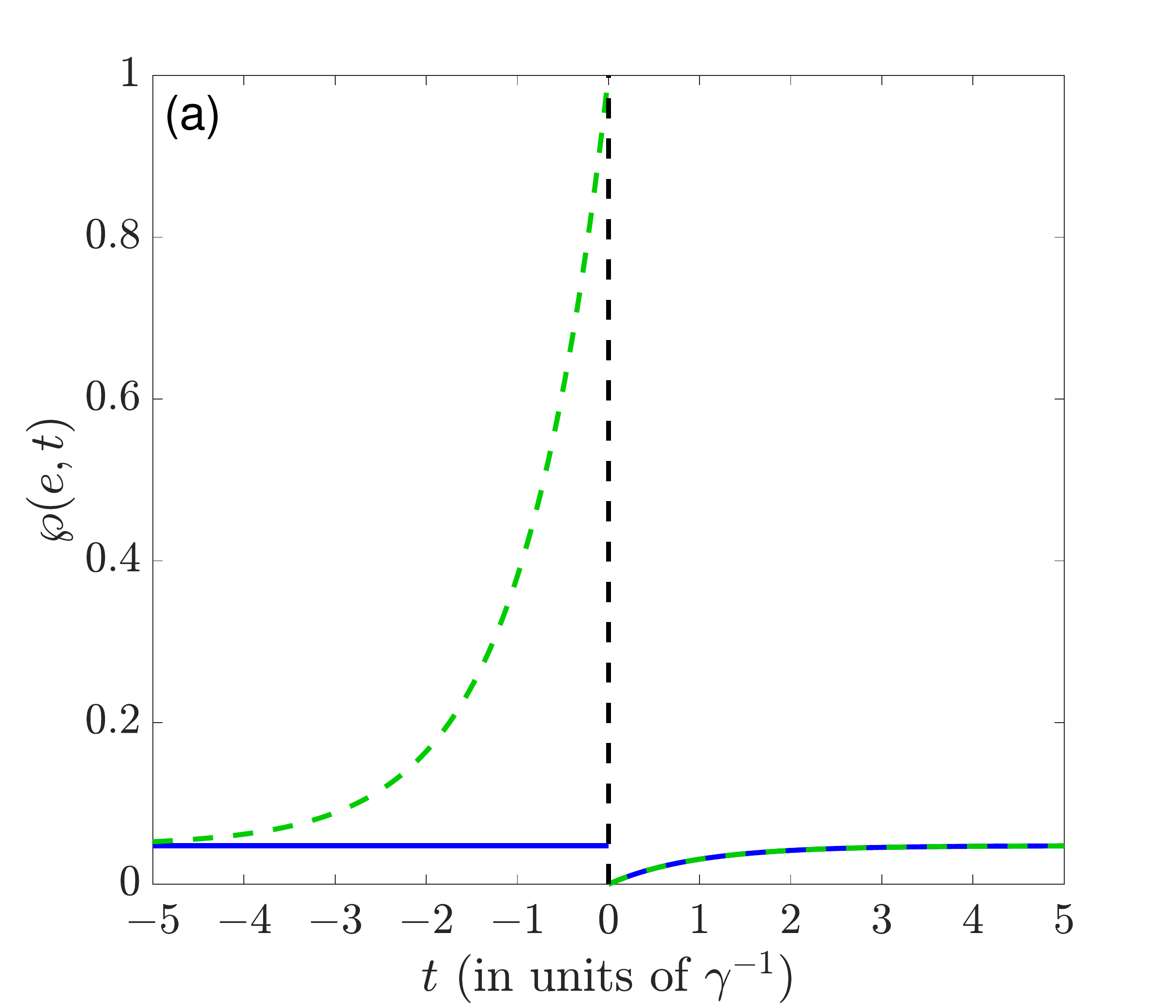}
\end{minipage}%
\begin{minipage}{0.3333\textwidth}
\includegraphics[scale = 0.25]{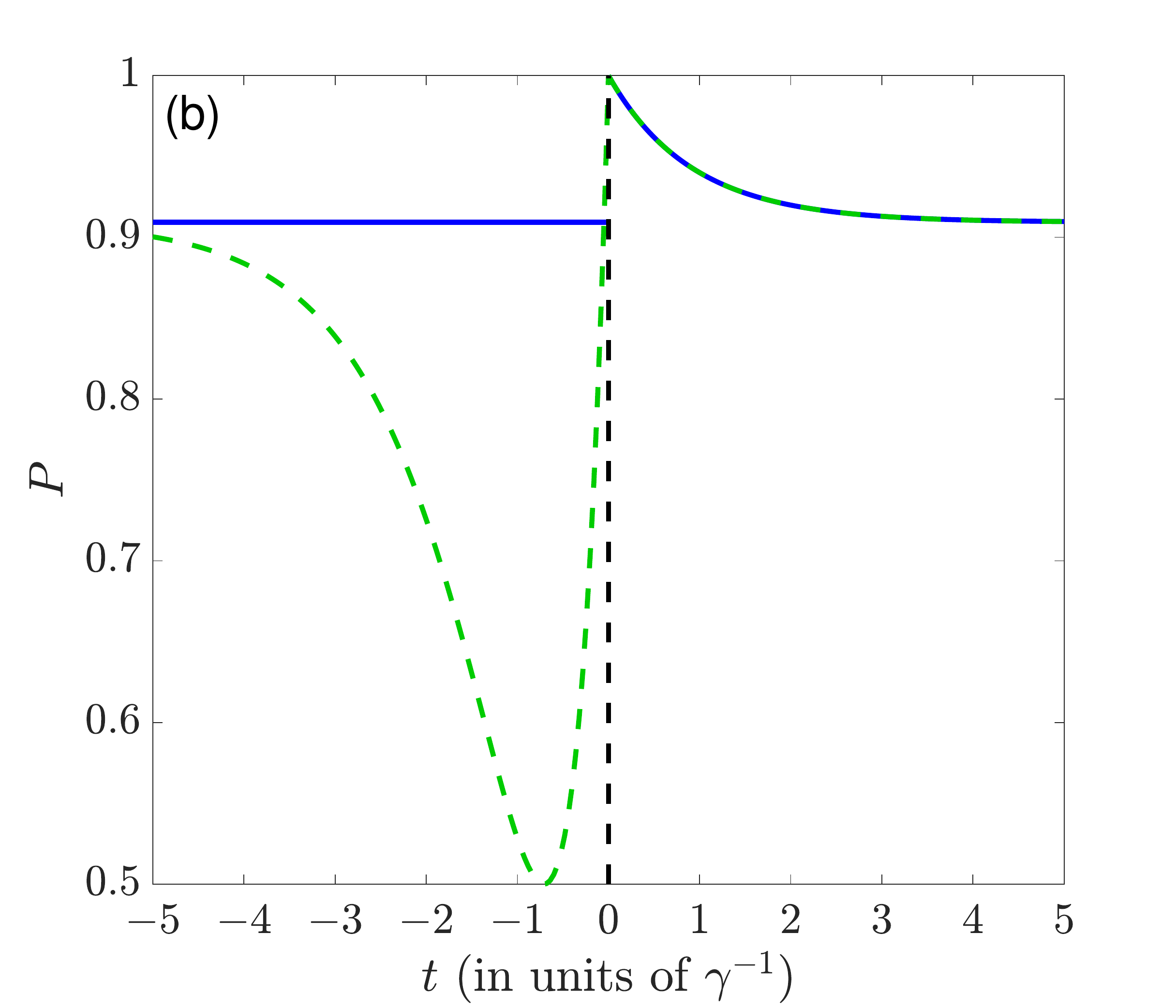}
\end{minipage}%
\begin{minipage}{0.3333\textwidth}
\includegraphics[scale = 0.25]{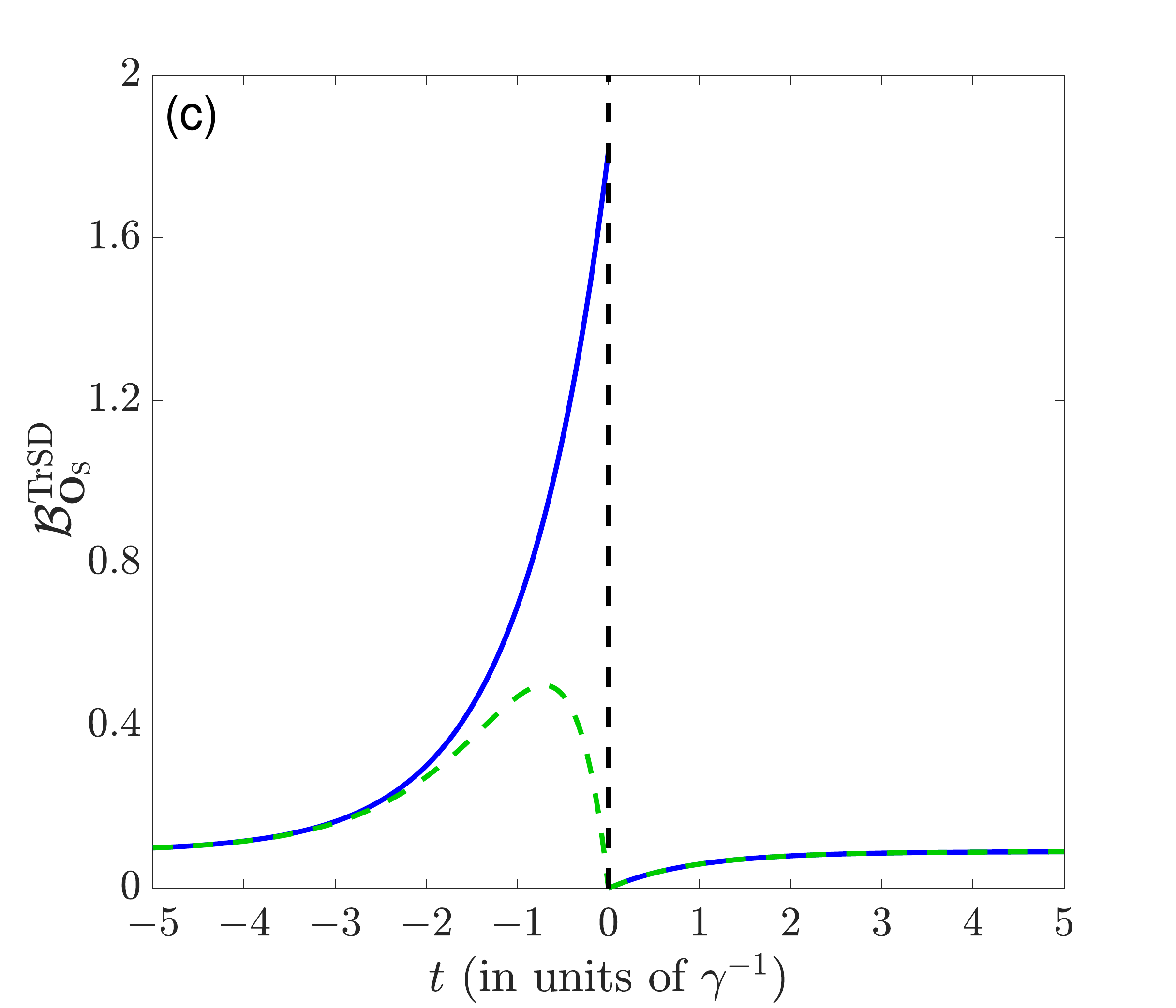}
\end{minipage}
\caption{(a) The filtered and smoothed probability of the system being in 
the excited state. Before the jump (at time $t = 0$) the smoothed state begins
to increase until it reaches unity right before the jump. 
(b) The purity for the filtered (blue line) and smoothed (green 
dashed line) states. The purity of the smoothed state begins to decrease 
before the jump due to the state having access to the future measurement 
record. (c) The past-future expected cost function for the filtered, using 
\erf{Fil-Sm-Risk}, and smoothed, using \erf{simp-risk}, states. The expected cost for the smoothed 
estimate of the state is seen to be a better estimator than the filtered 
estimate due to its smaller expected cost before the jump. In all cases, we 
have taken $\epsilon = 0.05\gamma$ and the limit $\delta \to 0^+$.}
\label{Fig-Class_Syst}
\end{figure*}

To gain some physical understanding of $\rho\sm\cl(t)$ in this system, let us 
compare it to the filtered state. We can see in 
Fig.~\ref{Fig-Class_Syst}~(a), 
that, unsurprisingly, prior to the jump occurring at $t=0$, 
the filtered state (solid blue line) remains in its steady state 
$\wp\fil\ss(e) = \epsilon/(\gamma + \epsilon)$ until the jump where upon it
is projected into the ground state. The smoothed quantum state (dashed 
green line), on the other hand, begins to diverge from the filtered state 
as the jump approaches and just prior to the jump reaches the excited state
before it is projected to the ground state. This divergence is expected as 
the smoothed state `knows' that a jump is about to occur, as it is 
conditioned on the future measurement record. Furthermore, since the true 
state can only be in either the ground or excited state, it means that for 
a jump to occur the system must have been in the excited state. 

If we look at the purity of the filtered and smoothed quantum state in 
Fig.~\ref{Fig-Class_Syst}~(b), we see that as the smoothed quantum state 
begins to deviate from the filtered state, the purity begins to drop 
rapidly. Such a drop is not surprising as in order for the smoothed state 
to reach the excited state it must pass through the maximally mixed state, 
resulting in the smoothed quantum state having minimal purity prior to the 
jump. However, this brings up an interesting point. 
From \erf{simp-risk}, we know that the expected cost function for the 
optimal state is equal to the average difference between the purity of the 
true state and the conditioned state. However, for this system, the true 
state is pure irrespective of the unobserved measurement record and 
\erf{simp-risk} reduces to the impurity of the conditioned state. This 
means that, since the purity of the smoothed state decreases below that of 
the filtered state just prior to the jump, the filtered state seemingly 
gives a lower expected cost function and would be the optimal estimator of 
the true state, not the smoothed quantum state. However, this is not true, 
as stated earlier and proved in Ref.~\cite{LGW-PRA21}. 

The issue lies in 
how the expected cost function is calculated for the filtered state. Using 
\erf{simp-risk} for the filtered state assumes that one only has access to 
Alice's past measurement record, whereas the smoothed quantum state assumes
that the past-future record is available. Thus, to compare the expected 
cost function of the filtered state to that of the smoothed state one must 
take the future measurement record into account and compute
\beq\label{Fil-Sm-Risk}
{\cal B}_{{\bf O}\sm}^{\rm TrSD}[\rho\fil] = 1 - 2\Tr[\rho\fil\rho\sm] + 
P(\rho\fil)\,.
\eeq
Note, this argument also holds for classical systems, with the appropriate 
analogues of the states and cost function. When computing the expected 
cost function of the filtered state in this way, we see, in 
Fig.~\ref{Fig-Class_Syst}~(c), that the smoothed quantum state has the 
lower expected cost function.\\

\section{Proof of Result 1: Bob uses X-Homodyne measurements}
\label{ssec-hom}
\begin{figure}[t!]
\includegraphics[scale=0.26]{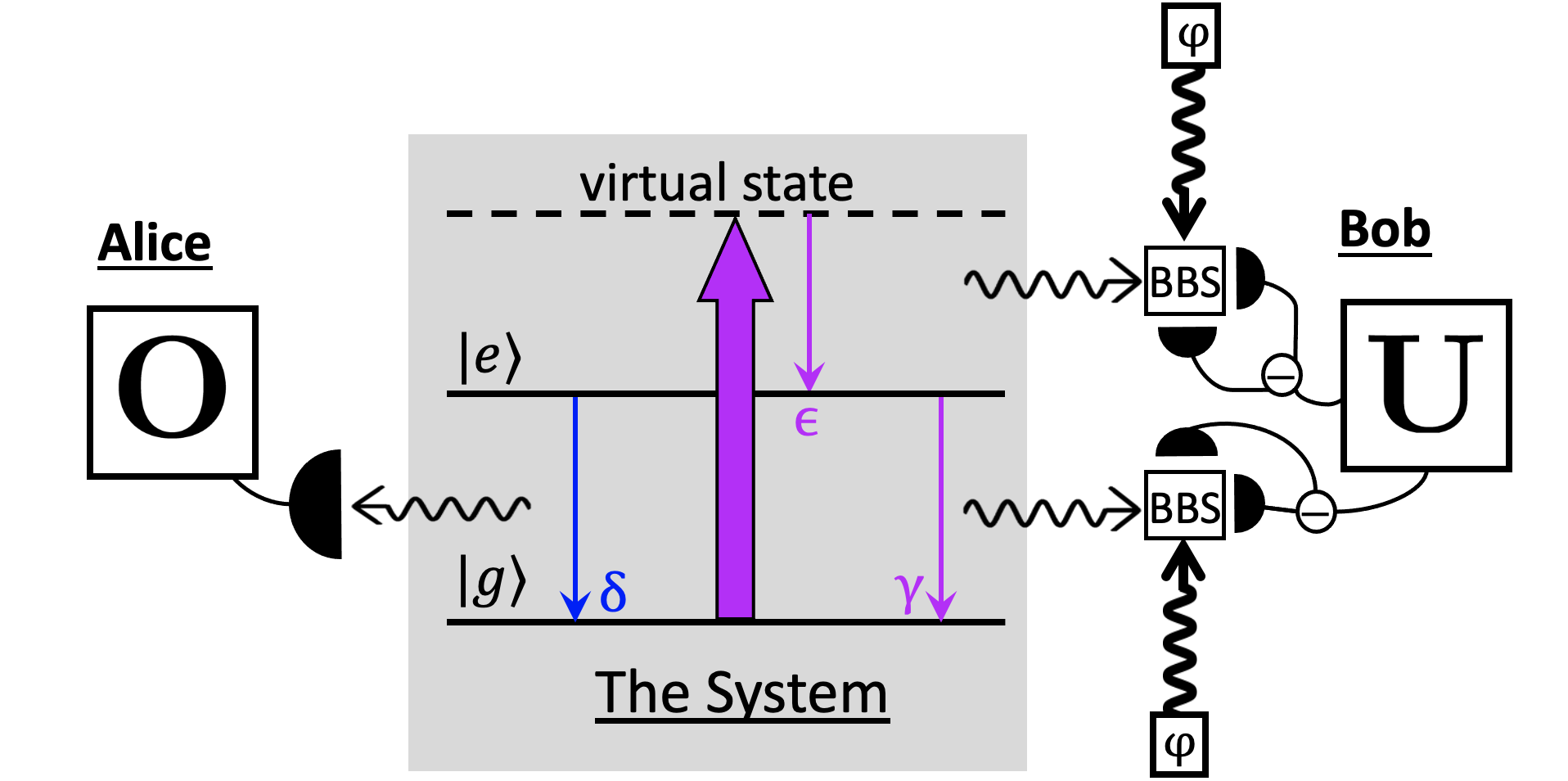}
\caption{The same physical system as in \frf{Fig-Photo} with Bob now 
measuring his $\gamma$ and $\epsilon$ channels both (independently) using X-homodyne measurements
(with local oscillator phase $\varphi = 0$). Here BBS stands for balanced beam splitter.}
\label{Fig-xhom}
\end{figure}

In this section we prove, by considering a different 
measurement choice for Bob, that the smoothed quantum state need not 
equal the classically smoothed quantum state. Here, instead of photon 
detection, Bob (perfectly) monitors the unobserved channels each with homodyne measurements. {\blk That is, the output light of the system is combined together with a local oscillator with phase $\phi$ on a 50:50 beam splitter, at which point, both outputs of the beam splitter are measured with photon detectors. The single two signals are then subtracted yielding quadrature information about the system. See Fig.~\ref{Fig-xhom}. Specifically, the measurement current obtained is \cite{WisMil10,GueWis20}
\beq
J_k\dd t = \ex{\hat{c}_{{\rm u},k}e^{i\phi_k} - \hat{c}_{{\rm u},k}\dg e^{-i\phi_k}}\dd t + \dd W_k\,,
\eeq
where $k = \gamma, \epsilon$, $\phi_k$ is the local oscillator phase for the $k$-channel, and $\dd W_k$ is the innovation, an infinitesimal Wiener increment satisfying 
\beq
\mathbb{E}\{\dd W_k\} = 0\,, \qquad \dd W_k \dd W_k'=\delta_{kk'}\dd t\,.
\eeq} 
Such a monitoring causes the 
true state to evolve according to the stochastic master equation \cite{WisMil10}
\beq\label{SME}
\begin{split}
\dd\rho\god =&\, {\cal G}[\hat{c}\ob]\rho\god\dd N\ob - 
\frac{1}{2}{\cal H}[\hat{c}\ob\dg\hat{c}\ob]\rho\god\dd t\\
&\qquad\qquad+ {\cal D}[\hat{\bf c}\un]\rho\god\dd t + 
{\cal H}[\blk\hat{\bf c}_\phi\dd {\bf W}\blk]\rho\god\,.
\end{split}
\eeq
{\blk Here $\hat{\bf c}_\phi = [\hat{c}_{{\rm u},\gamma} e^{i\phi_\gamma}, \hat{c}_{{\rm u},\epsilon} e^{i\phi_\epsilon}]$ and $\dd{\bf W} = [\dd W_\gamma, \dd W_\epsilon]\tp$. From this point forward, we will consider the case where $\phi_\gamma = \phi_\epsilon = 0$, \ie, X-homodyne measurements on both channels.} Note, this is called an X-homodyne measurement since the 
measurement current only contains information about the $x$-component of 
the Bloch vector ${\bf r} = (x,y,z)\tp$ that characterizes the state of the
qubit, where $\rho = \half(\hat{1} + {\bf r}\cdot\hat{\bm\sigma})$ and 
$\hat{\bm\sigma} = (\s{x},\s{y},\s{z})\tp$. 
The analysis that is to 
follow would also hold, in this system, for any choice of homodyne phase $\phi_\gamma = \phi_\epsilon = \phi$. 

The important part about this measurement scheme for Bob is that, between 
two observed jumps, $\rho\god$ is not restricted to the set 
$\{\ket{g}\bra{g},\ket{e}\bra{e}\}$, but instead can be in any pure state 
on the $x$-$z$ great circle of the Bloch sphere. The true 
state is pure between jumps because it starts in the 
(pure) ground state and remains pure since the system is perfectly 
monitored by both Alice and Bob. 
As for why the pure state can be  
confined to the $x$-$z$ great circle of the Bloch sphere, this is because, 
without Bob's measurement, the quantum state is confined to 
the $z$-axis of the Bloch sphere and conditioning on an X-homodyne 
measurement will only give information about the $x$-component of the 
Bloch vector. Hence, the 
true state will, typically, have a non-zero $x$-component, \blk whilst \blk the 
$y$-component will remain zero. 

To prove this more rigorously, one can obtain the 
stochastic differential equations for the Bloch vector of the true state 
from \erf{SME} subject to the initial condition $\rho\god(t_1^+) = 
\ket{g}\bra{g}$ and a no-jump record $\dd N\ob(t) = 0$. We will again 
take the limit $\delta \to 0^+$ to simplify the computation. For this 
measurement scenario we have
\begin{align}
\dd z &= \left[-\gamma(1 + z) + \epsilon(1-z)\right]\dd t \nn\\ & 
\qquad\qquad- \sqrt{\gamma}x(1+z)\dd W_\gamma 
+ \sqrt{\epsilon}x(1-z)\dd W_\epsilon\,,\\
\dd y &= y\left[-\frac{1}{2}(\gamma + \epsilon)\dd t - 
\sqrt{\gamma}x\dd W_\gamma - \sqrt{\epsilon}x\dd W_\epsilon\right]\,,\\
\dd x &= -\frac{1}{2}(\gamma + \epsilon) x\dd t + \sqrt{\gamma}
(1 + z - x^2)\dd W_\gamma \\
&\qquad\qquad\qquad\qquad\qquad + \sqrt{\epsilon}(1 - z - x^2)
\dd W_\epsilon\,,
\end{align}
with $z(t_1^+) = -1$ and $y(t_1^+) = x(t_1^+) = 0$. 
With these equations we see that, due to the initial condition, the $y$-component will
remain zero until the next observed jump. 

Since the true state is restricted to the $x$-$z$ great circle, we can 
reparametrize the true state by the angle $\theta$ of $y$-rotation from the
positive $z$-axis instead of the Bloch vector. That is, defining  
\beq
\rho\god(\theta;t) = \half(\hat{1} + \sin\theta(t)\s{x} + 
\cos\theta(t)\s{z})\,,
\eeq
we have ${\mathbb T}_t = \{\rho\god(\theta;t):\theta \in [0, 2\pi)\}$ for 
all $t \in (t_1,t_2)$. 
With this $\theta$-parametrization we can reduce the evolution of the 
true state between two observed jumps to a single stochastic differential 
equation for $\theta$. Using the \ito~formula~\cite{GardinerBook,WisMil10} 
to move from a differential equation in $\cos\theta$ (or $\sin\theta$) to 
one in $\theta$, we obtain
\beq\label{Langevin-Eq}
\dd \theta = A(\theta)\dd t + B_\gamma(\theta)\dd W_\gamma + 
B_\epsilon(\theta)\dd W_\epsilon\,.
\eeq
Here, $A(\theta) = \sin\theta[\half(\gamma + \epsilon)\cos\theta + 
(\gamma - \epsilon)]$, $B_\gamma(\theta) = \sqrt{\gamma}(1 + \cos\theta)$ 
and $B_\epsilon(\theta) = -\sqrt{\epsilon}(1 - \cos\theta)$. 

\begin{figure}[t!]
\includegraphics[scale=0.28]{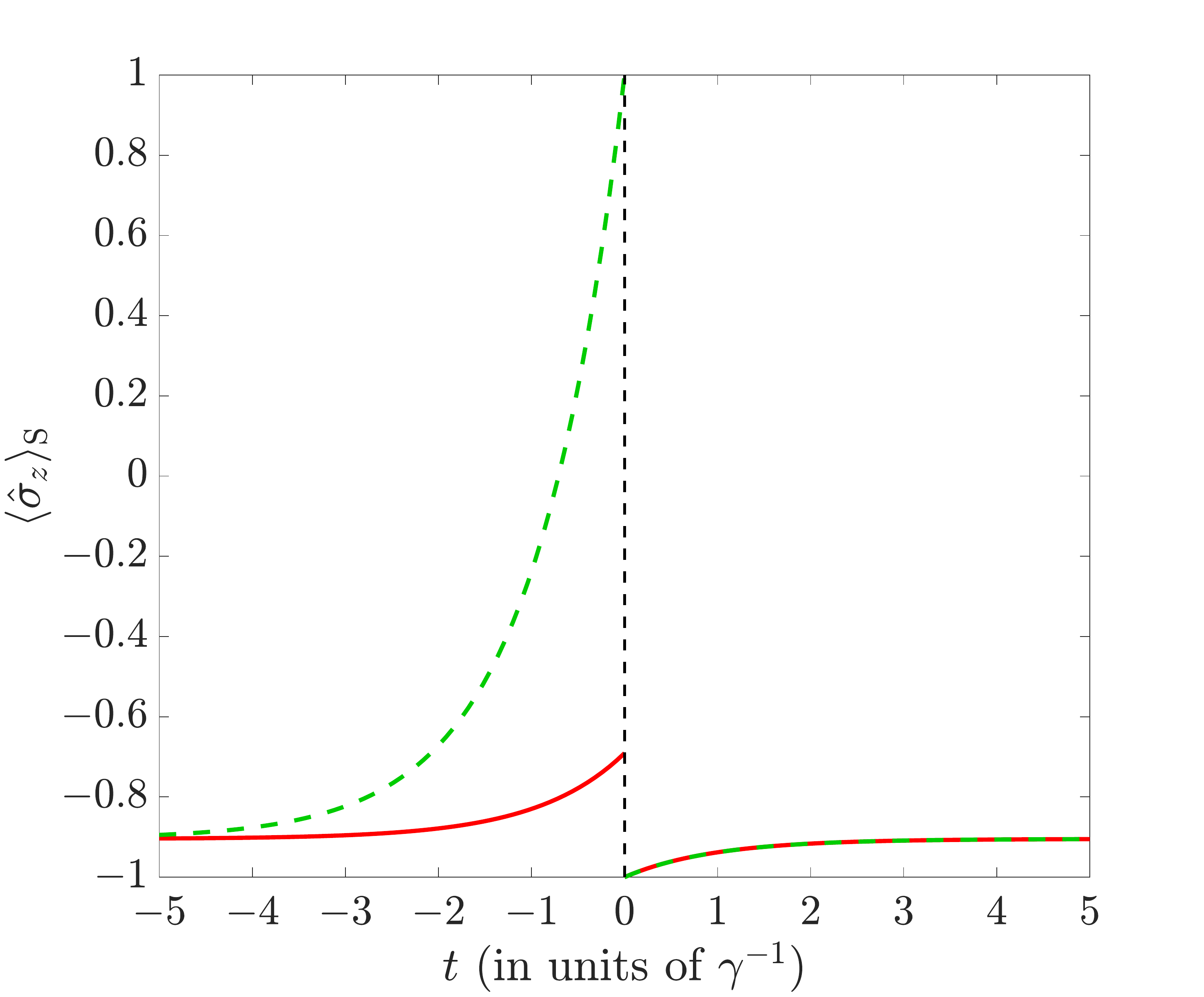}
\caption{Comparing the conditional average of the $z$-component of the 
Bloch vector when Alice's measurement record is obtained by photon 
detection and Bob's measurement record is obtained by either photon 
detection (green dashed line) and an X-homodyne (red line) measurement. 
Here, we have taken $\delta \to 0^+$ and $\epsilon = 0.05\gamma$ and an 
observed detection (jump) occurs at time $t=0$. The $z$-component for the
smoothed state using an X-homodyne measurement for Bob differs from the 
classically smoothed state despite the filtered state being diagonal in 
the $\s{z}$-basis, proving result 1.}
\label{Fig-Hyp1}
\end{figure}

We can now begin to compute the smoothed quantum state for this case. From 
\erf{QSS}, we have
\beq\label{qsm-int}
\begin{split}
\rho\sm(t) &= \int_{0}^{2\pi} \wp(\theta;t|\both{\bf O})\rho\god(\theta) 
\dd\theta\\
&\propto \int_{0}^{2\pi} \Tr[\hat{E}\rfil(t)\rho\god(\theta)] 
\wp(\theta;t|\past{\bf O}_t) \rho\god(\theta)\dd\theta\,.
\end{split}
\eeq 
As we are only considering the evolution between two observed jumps, we 
only need to find $\wp(\theta;t|\past{\bf O}_t) = \wp(\theta;t|{\rm N.J.})$, 
as the retrofiltered effect can be computed via \erf{Erfil}. Here the 
conditioning ${\rm N.J.}$ stands for a no-jump record.
Since \erf{Langevin-Eq} is in the form of a Langevin equation, it can be 
mapped to a Fokker-Planck equation~\cite{GardinerBook,GamWis05,WisMil10,CGLW21}
describing the evolution of the probability density of $\theta$ for unknown
Wiener processes. However, since \erf{Langevin-Eq} assumes a no-jump 
observed record, the probability density the Fokker-Planck equation  
describes is in fact $\wp(\theta;t|{\rm N.J.})$. For this Langevin equation, 
the corresponding Fokker-Planck equation is
\beq
\begin{split}\label{FPE}
\partial_t \wp(\theta;t|{\rm N.J.}) =& -\partial_\theta 
A(\theta)\wp(\theta;t|{\rm N.J.}) \\ 
&\,\,\,\, + \frac{1}{2}\sum_{k \in 
\{\gamma,\epsilon\}}\partial_\theta^2 B_k(\theta)^2 
\wp(\theta;t|{\rm N.J.})\,,
\end{split}
\eeq
where $\partial_x = \partial/\partial x$, with the initial condition 
$\wp(\theta; t_1^+|{\rm N.J.}) = \delta(\theta - \pi)$ corresponding to the 
ground state and the boundary condition $\wp(0;t|{\rm N.J.}) = \wp(2\pi; 
t|{\rm N.J.})$. We solve this Fokker-Planck equation 
numerically using Mathematica's NDSolve function with a Gaussian initial 
condition $g(\theta;\mu = \pi,V = 0.01\gamma)\approx\delta(\theta - \pi)$, 
where the mean of the Gaussian is $\mu$ and variance $V$. With the 
probability density found, we can now compute the smoothed quantum state in
the X-homodyne case for comparison to the photon detection case.

As an aside, in general, to calculate the smoothed quantum state 
requires using an ensemble of unnormalized true states generated according 
to an ostensible probability distribution. See, for example, Refs.~\cite{GueWis20,CGLW21}. 
That is, one must calculate an extra stochastic variable, the norm of each possible true state. 
This applies, in general, even in the current simple system where we 
can generate the ensemble of true states by solving a Fokker-Planck equation. 
Specifically, the Fokker-Planck equation must be modified to describe the joint 
probability of $\theta$ and the normalization~\cite{GamWis05,CGLW21}. 
In the present case, however, since we are considering the limit 
$\delta \to 0^+$, the amount of information Alice gains from a 
no-detection event is negligible, as this is almost always the result, 
causing the equation of the true state to reduce to just the unobserved 
evolution. Since the actual probabilities, for this case, can be computed 
easily from the distributions for $\dd W_\gamma$ and $\dd W_\epsilon$, 
the normalized equation for the true state can be used.  

We can now begin to compute the smoothed quantum state. 
%
%
We can see that, in this homodyne case, the smoothed quantum
state will be diagonal in the $\s{z}$-basis because of a symmetry in the dynamics. 
Specifically, since there are no unitary dynamics driving the system in a 
particular way, for any unobserved measurement record that  
causes the true state to rotate clockwise on the $x$-$z$ great 
circle, there is an equally likely record of the opposite sign that causes the state 
to evolve in exactly the same way in the counter-clockwise direction. 
Importantly, it is equally likely even given the future record (the jump) that 
Alice observes because the excited state probability is the same for both directions 
of rotation. 
Thus, when Alice averages over the 
possible unobserved records, each true state can be paired with its mirror 
image about the $z$-axis, cancelling the $x$-component of the Bloch vector. 

We can thus easily compare the 
X-homodyne smoothed quantum state with the classically smoothed quantum 
state by only looking at their $z$-components. 
As \frf{Fig-Hyp1} shows, there is  a clear difference between them.  (As explained 
in Sec.~\ref{ssec-photo}, in the limit $\delta \to 0^+$ we need be concerned only with the smoothed 
quantum state in a time of order $1/(\gamma + \epsilon)$ before the final jump.) 
Thus, by this example, we have proven our first result: the commuting of the filtered 
state and retrofiltered effect is {\em not} 
sufficient for $\rho\sm(t)$ to equal $\rho\sm\cl(t)$. 

In this example, the obvious difference between the smoothed state obtained classically 
(when Alice assumes an orthogonal basis for the true state) and that obtained when Bob performs a 
homodyne measurement (when Alice takes this into account) is the value of $z$ for the smoothed 
state immediately prior to the jump. We can intuitively understand this as follows. 
Since the true state in the latter case can be in a 
superposition of both the ground and excited state, as opposed to just 
being in one of these states, a jump can occur even when the system is not 
in the excited state. Thus, when Alice is estimating the true state using 
smoothing, she cannot be certain that the true state was in the excited 
state just prior to the transition, unlike in the classical (photon detection) case 
where she is certain. As such, \blk her smoothed estimate in the homodyne case \blk 
only moves somewhat closer to the excited state as the jump approaches. 


\section{Proof of Result 2: Bob uses an adaptive measurement}
\label{ssec-adap}

In the preceding section, the non-classical smoothed state was still 
diagonal in the same basis as the classical smoothed state, and therefore 
diagonal in the same basis as the filtered state and retrofiltered effect  
(whose co-diagonality defines the scenario we are investigating).
In this section we show the stronger 
result that the smoothed quantum state is not necessarily even 
co-diagonal with $\rho\fil$ and $\hat E\rfil$. 
As we saw in the X-homodyne example, 
the smoothed quantum state was diagonal in the $\s{z}$-basis because 
Bob's measurement gave the set of possible true states a symmetry about the
$z$-axis of the Bloch sphere. To avoid reasoning of this sort for this 
final case, Bob's measurement is chosen to break this symmetry in the set 
of possible true states. 

To achieve an asymmetric distribution of true states, we allow Bob to 
use an adaptive measurement 
involving finite strength local oscillators on the 
unobserved channels; see \frf{Fig-adap}. Here, ``finite strength'' 
means that the local oscillator intensity is comparable to the intensity of 
light emitted by the system, so the detection still resolves individual photons 
(unlike the strong local oscillator case of a homodyne measurement). 
The measurement is ``adaptive'' in the sense that  after 
every detection event, the amplitude (strength and phase) of the local oscillators 
can be changed, depending on the current settings of these amplitudes, and the 
type of photodection that occurs (if more than one detector is used, which is the 
case for our system here, with two unobserved channels). Measurements of this kind 
have been studied theoretically for many decades~\cite{Dol73,Hel76,WisToo99,KarWis11,WisGam12,DarWis14,WarWis19,WarWis19q}. 

A non-trivial property of adaptive measurements of the sort described is that they can 
{\blk constrain} the stochastic evolution of the true state of the system to a finite and time-independent 
set ${\mathbb T}$. \blk Note the lack of a subscript $t$\blk. Together with the corresponding 
stationary probabilities of each state in ${\mathbb T}$, 
this comprises a so-called {\em physically realizable ensemble} (PRE)~\cite{WisVac01,KarWis11}. 
The significance of 
``physically realizable'' here is as follows. Consider a Markovian 
(in the strongest sense~\cite{LiHallWis18}) open quantum system whose 
unconditional evolution is described by a Lindblad master equation, where, 
in the long-time limit, $t \geq t\ss$, the system reaches a unique stationary 
solution that is mixed. Owing to the fact that a mixed quantum state can be 
decomposed into a weighted ensemble of pure states in
infinitely many ways, there are different interpretations of {\blk how} the underlying 
pure state dynamics of the system unfold. However, only some 
of these pure state 
ensembles are physically realizable, meaning that there exists a way to continuously monitor the environment (without affecting 
the unconditional evolution) such that the conditioned state at times after 
$t\ss$ is confined to ${\mathbb T}$  with the corresponding probabilities being 
realized in the ergodic sense. 

The classical ensemble of Sec.~\ref{ssec-photo}, with ${\mathbb T} = \{\ket{g}\bra{g},\ket{e}\bra{e} \}$, is an example of such a PRE, 
but in general the states in ${\mathbb T}$ need not be orthogonal~\cite{WisToo99,KarWis11,DarWis14,WarWis19,WarWis19q}. 
This is essential for finding an asymmetric (under application of $\s{z}$) steady-state ensemble. 
(In our case, this steady state means a long time, compared to $1/(\gamma + \epsilon)$, 
after Alice's last jump, which is always the limit we can consider when Alice's jump rate $\delta \to 0^+$; 
see more below.) For our 
system, the particular adaptive measurement scheme is chosen so that
${\mathbb T}$ comprises three states, with no symmetry under application of $\s{z}$. The 
conditioned dynamics causes the true state to cyclically transition 
between these states. This is possible only in certain parameter regimes, 
which is why we chose $\epsilon = 0.05\gamma$. 

\begin{figure}[t!]
\includegraphics[scale=0.26]{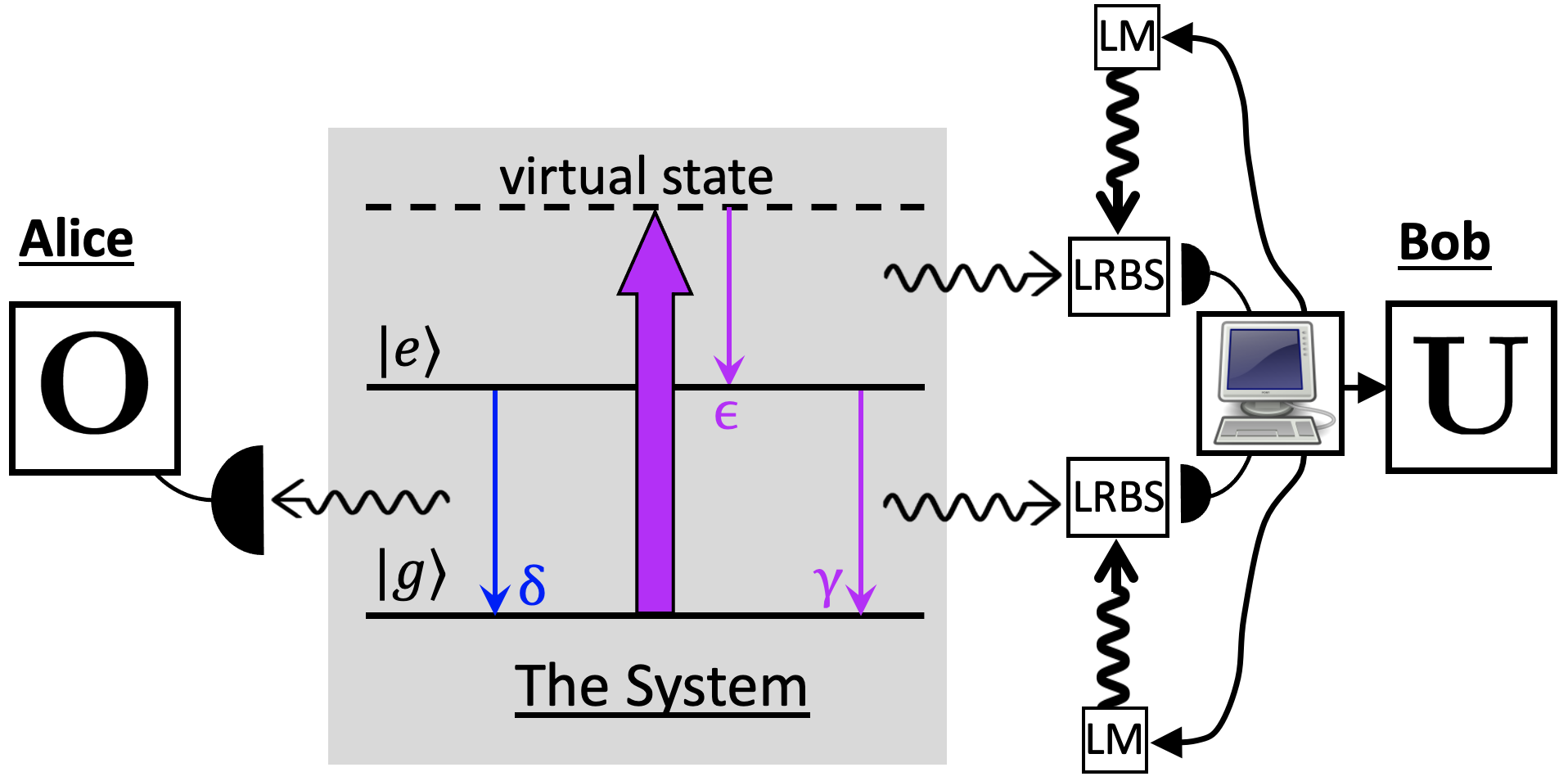}
\caption{The same physical system in \frf{Fig-Photo} with Bob using the 
adaptive measurement scheme in App.~\ref{App:measScheme}. In this figure, LRBS stands for low-reflectivity beam splitter and LM stands for light modulator.}
\label{Fig-adap}
\end{figure}

{\blk A detailed discussion of PREs in the context of the photon emission and absorption master equation is given in Refs.~\cite{WarWis19}~\S4.4.2.}
\blk In our case, the scenario is modified slightly as it is Bob's measurement record which causes Alice's filtered state (as 
opposed to the unconditioned state) to become pure upon conditioning. That is, we make the substitution $\rho \to \rho\fil$ and it is Bob's measurement 
that causes the true state to become pure when deriving the PRE for the system. Note, \blk there is a subtlety in regards to the \blk filtered state 
that one should be aware of when making this substitution. In the standard PRE scenario, it is necessary for the unconditioned state to have reached a 
unique stationary solution. However, the filtered state is a stochastic quantity and in general will not reach a unique stationary solution in the long-time limit. As such, there are only select cases, i.e., when the filtered state (or some deterministic property of the state \cite{LCW-QS21}) reaches a unique steady state, that we can apply the PRE theory in this manner. 

In the example that we consider, such a steady-state will usually exist for the filtered state, provided the time between consecutive jumps, $\tau$, is 
typically much longer than the time in which the system equilibriates. More formally, when 
$\tau = (\delta\bra{e}\rho\ket{e})^{-1} \gg (\gamma + \epsilon + \delta)^{-1}$. Since, after the first jump, $\rho(t_1) = \ket{g}\bra{g}$, it will 
always be the case that $\bra{e}\rho\ket{e} \leq \bra{e}\rho\ss\ket{e} = \epsilon(\gamma + \epsilon + \delta)\inv$. Thus, the filtered state is likely to reach steady state between consecutive jumps if we operate in the parameter regime
\beq
\delta\epsilon \ll (\gamma + \epsilon + \delta)^2 \approx (\gamma + \epsilon)^2\,,
\eeq 
where, for the final approximation, we have assumed $\delta \ll \gamma + \epsilon$. Note, for the parameter regime we have been considering thus far 
($\epsilon = 0.05\gamma$), we require that $\delta \ll 22\gamma$. As has been the case for the other measurement scenarios, we will, for simplicity, take the limit $\delta\to 0^+$.\blk

For the stationary solution of the filtered state in this example,
\beq
\rho\fil\ss = \frac{\epsilon}{\gamma + \epsilon}\ket{e}\bra{e} + 
\frac{\gamma}{\gamma + \epsilon}\ket{g}\bra{g}\,,
\eeq
the cyclic physically realizable ensemble 
$\{\rho(\theta), \wp_\theta\}_{\theta \in \{\alpha,\beta,\phi\}}$ chosen, from the 
possible valid ensembles, is displayed in \frf{Simp_Bloch} with the angles and corresponding probability weights. 
It should be 
emphasized that the implemented measurement strategy~\cite{WarWis19} only 
causes the true state to undergo the cyclical dynamics depicted in 
\frf{Simp_Bloch} when the filtered state is in steady state. Outside of 
this regime the dynamics of the true state may be more complicated, but
when averaged still result in the transient dynamics of the filtered state. {\blk The reader is referred to App.~\ref{App:measScheme} for the local oscillator settings that achieve the PRE shown in \frf{Simp_Bloch}.}

\begin{figure}[t!]
\includegraphics[scale = 0.33]{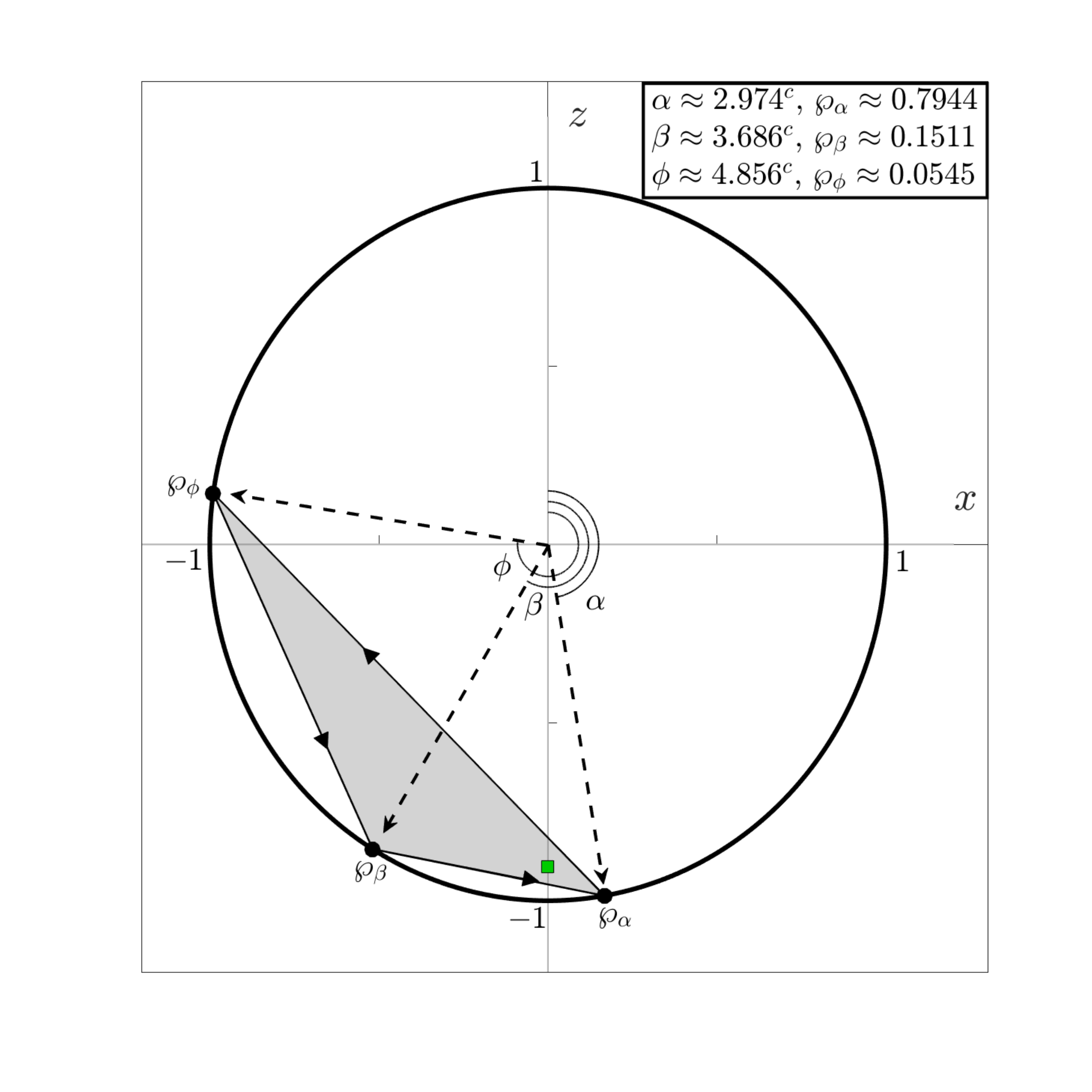}
\caption{\blk The $x$-$z$ great circle of the qubit's Bloch sphere showing\blk~ the cyclic physically realizable ensemble chosen for the example 
qubit system subject to the measurement scheme in \frf{Fig-adap} plotted on
the $x$-$z$ great circle. The three black points on the circle are the 
three pure states in the ensemble \blk with there corresponding angles and occupation 
probabilities given in the top right corner \blk
(see Ref.~\cite{WarWis19} for how these 
are calculated) and the arrows connecting them shows the cyclic dynamics 
that results from the adaptive measurement strategy. The shaded region is 
where both the filtered and smoothed quantum state must reside at times 
after $t\ss$. Lastly, the square marker indicates $\rho\fil\ss$. 
See \frf{Fig-Hyp1} for the parameter details. }
\label{Simp_Bloch}
\end{figure}

With the measurement scheme and dynamics of the true quantum state covered 
we can now begin to compute the smoothed quantum state for this case. 
As in the previous two cases, the 
smoothed quantum state differs from the filtered state only for a 
time of order $1/(\gamma+\epsilon)$ 
prior to the second observed jump. 
As such we only need to compute the smoothed quantum state when the 
retrofiltered effect is outside of steady state. Since we are working in 
the limit as $\delta\to 0^+$, it will typically (in the strict sense) be the case that enough 
time has passed for the filtered state to have reached steady state well 
before the next jump. As such, over the time prior to the second 
jump that we are interested in, the true state will be cyclically jumping 
between the three pure states in the PRE. Thus the smoothed quantum state 
over this time region can be computed via
\beq
\rho\sm(t) \propto \sum_{\theta \in\{\alpha,\beta,\gamma\}} \Tr\left[\hat{E}\rfil(t)\rho(\theta)\right]\wp(\theta;t|\past{\bf O}_t)\rho(\theta)\,.
\eeq
All that remains to compute the smoothed quantum state is to determine 
the probability distribution $\wp(x|\past{\bf O}_t)$. This distribution 
is, by definition, the occupation probabilities of the PRE states, 
i.e., $\wp(\theta;t|\past{\bf O}) = \wp_\theta$.

\begin{figure}[t!]
\includegraphics[scale=0.28]{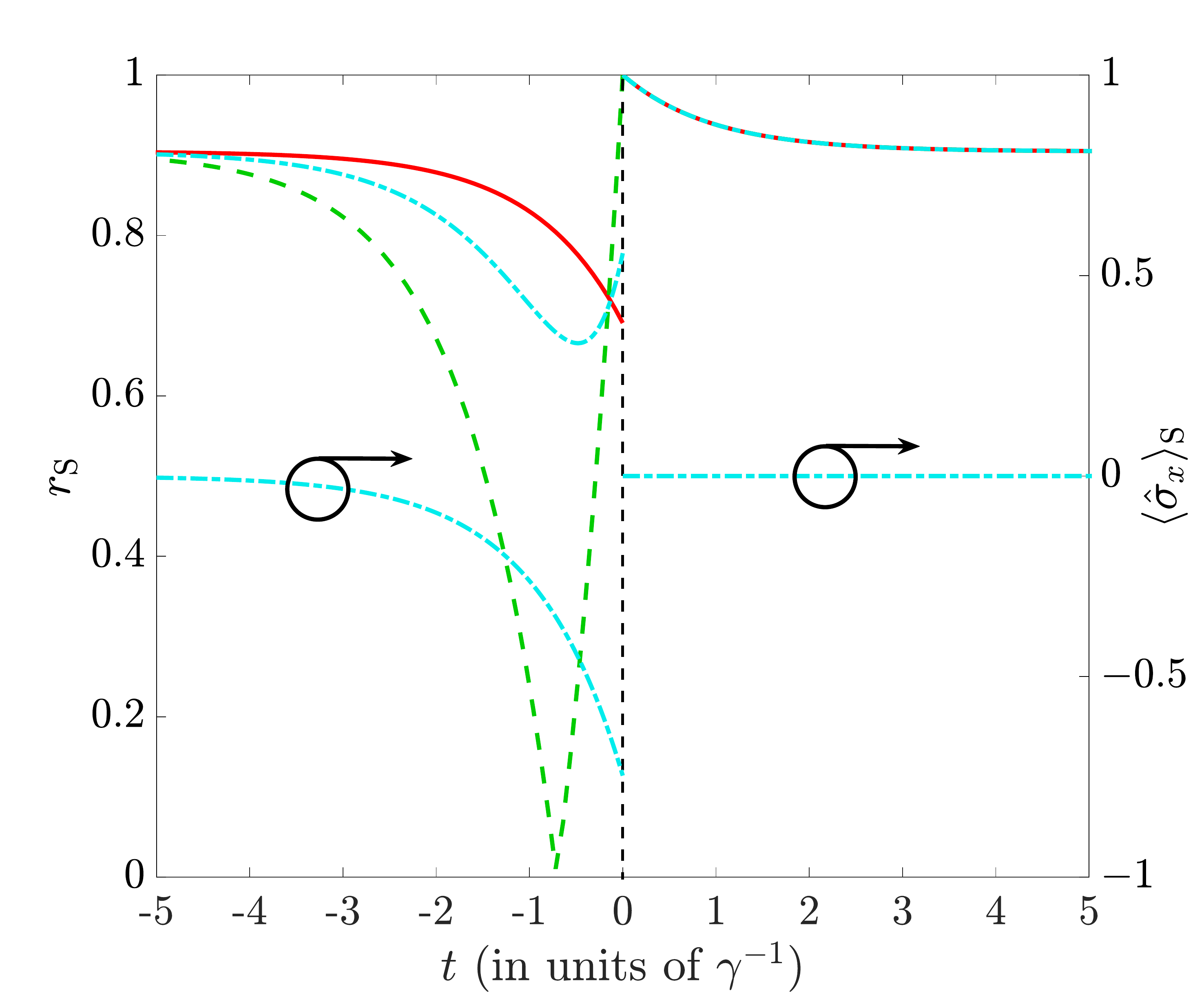}
\caption{The length (left-hand-side axis) and the $x$-component 
(right-hand-side axis) of the Bloch vector for the smoothed quantum state 
when Alice's measurement record is obtained by photon detection and Bob's 
record by either photodetection (green dashed line), X-homodyne (red solid 
line), or an adaptive weak local oscillator (cyan dot-dashed line). The 
$x$-component of the smoothed state when Bob implements photodetection or an
X-homodyne measurement are not shown as they are zero at all times.}

\label{Fig-Hyp3}
\end{figure}
\begin{figure*}[t!]
\begin{minipage}{0.5\textwidth}
\includegraphics[scale = 0.365]{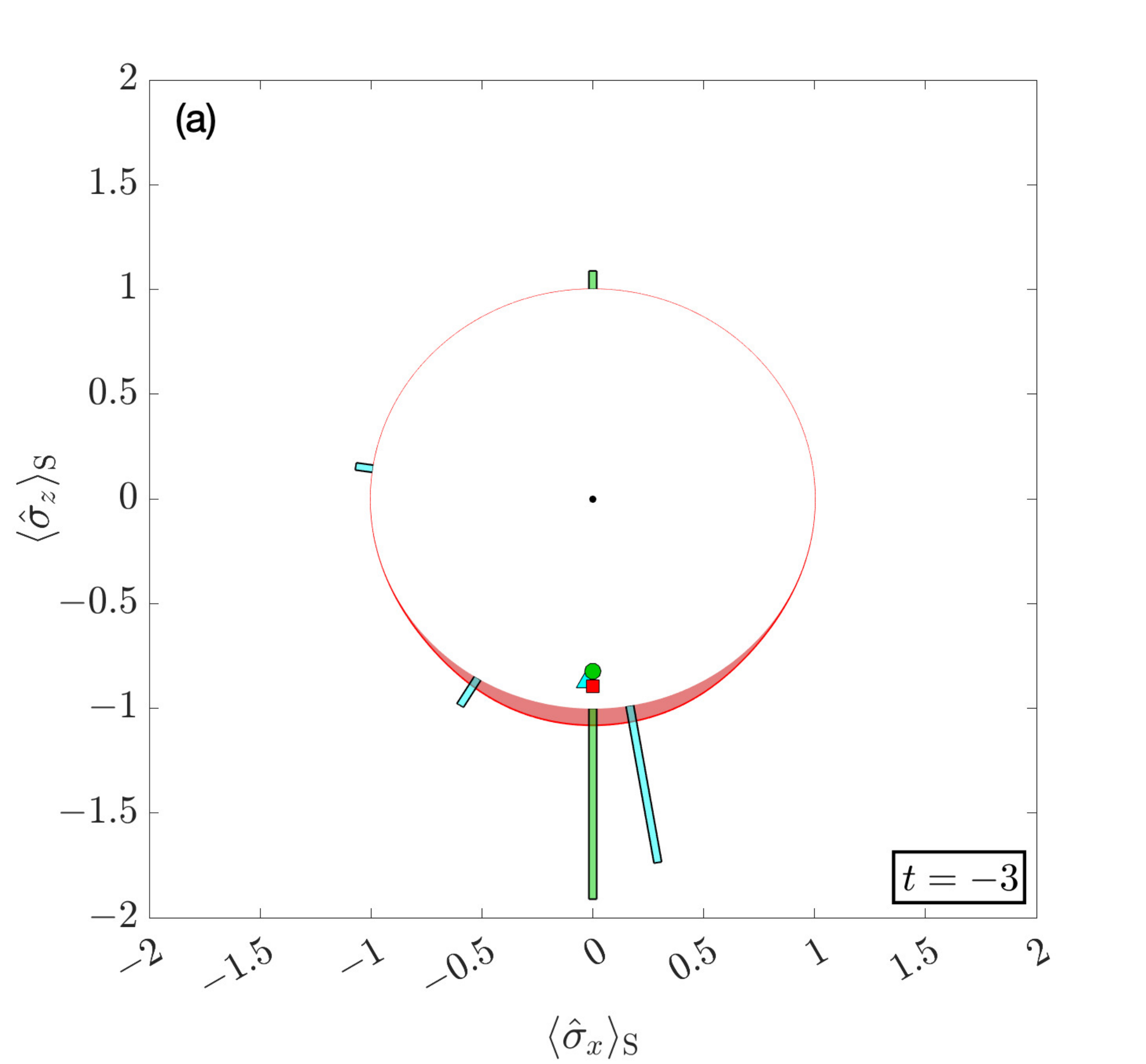}
\end{minipage}%
\begin{minipage}{0.5\textwidth}
\includegraphics[scale = 0.365]{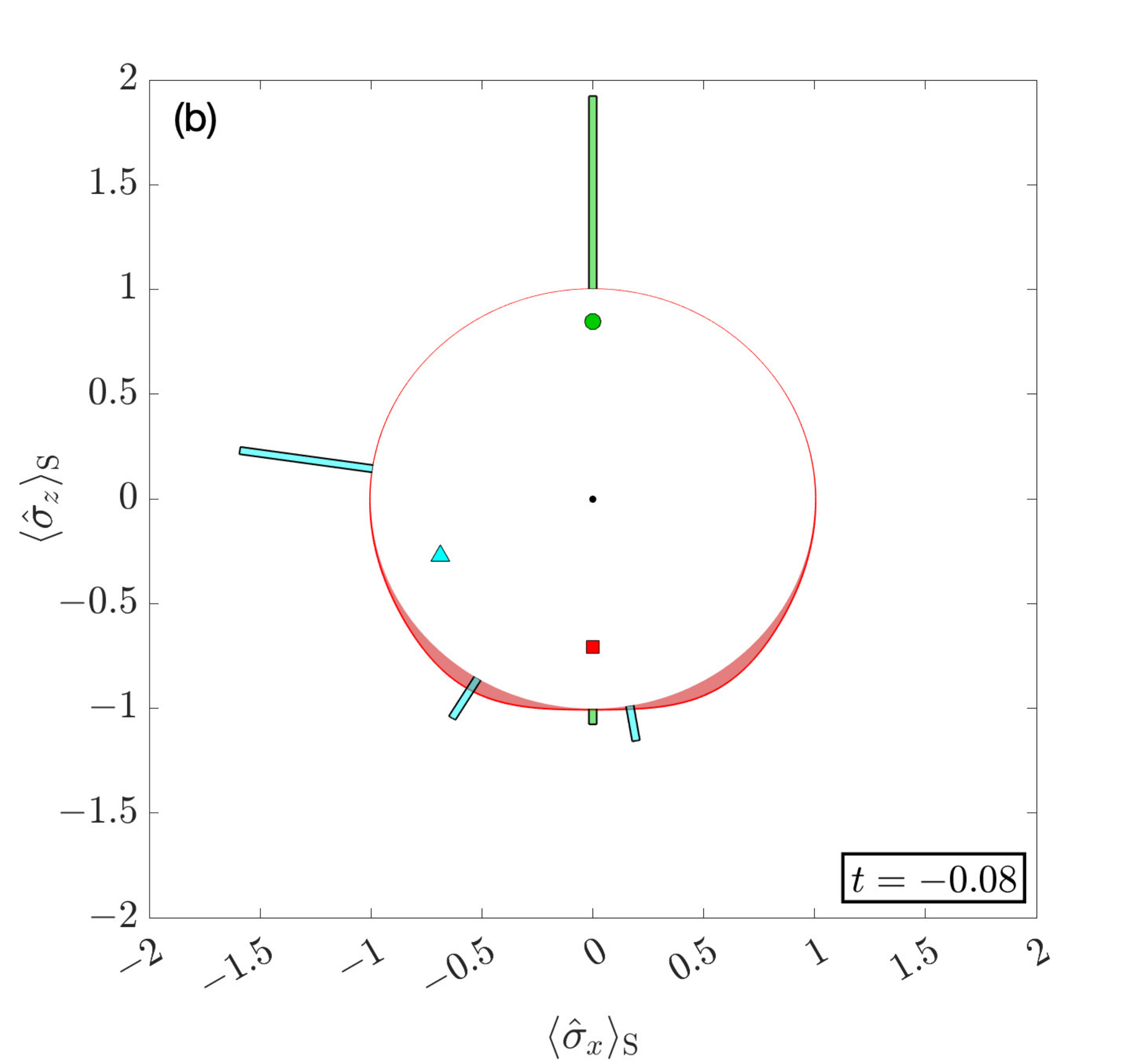}
\end{minipage}%
\vspace{-2mm}
\caption{(a) and (b) show the $x$-$z$ great circles of the Bloch sphere plotted at two 
different times prior to the observed jump. On the surface of the circle we
have plotted the probability distributions for the case where Bob measures
the environment using photodetection (green), an X-homodyne (red), and an 
adaptive scheme (cyan) (all of which have a zero $y$-component). The 
relative area of the bars indicates the occupation probability for
that state. Note, the total area of both the the
green and cyan bars are equal but the area under the red curve has been 
scaled by a factor of $3$ for clarity. The markers (circle, square and triangle) on the interior correspond to the smoothed mean using the 
correspondingly coloured distribution. See \frf{Fig-Hyp1} for the parameter
details.}
\label{daryPRE}
\end{figure*}

To check whether the smoothed quantum state is non-diagonal in the 
$\s{z}$-basis, we only need to look at the $x$-component of its Bloch 
vector, as the $y$-component is always zero for this adaptive local oscillator scheme. In 
\frf{Fig-Hyp3}, the $x$-component (right-hand-axis) and the length of the 
Bloch vector (left-hand-axis), defined as 
$r = \sqrt{\ex{\s{x}}^2 + \ex{\s{y}}^2 + \ex{\s{z}}^2}$, for the smoothed 
quantum state are shown for this scheme (cyan). 
There are a couple of interesting features in 
this example, the most important being that the Bloch vector of the 
smoothed quantum state prior to the jump has non-zero $x$-component. Thus, 
the smoothed quantum state is {\it not} diagonal in the $\s{z}$-basis, the 
shared basis of the filtered state and retrofiltered effect, proving our 
second result.

The reason the smoothed quantum state becomes non-diagonal 
is that Alice is able to infer from $\fut{\bf O}_t$, 
specifically the upcoming jump, that the two states in the PRE with a 
negative $x$-component are more likely to have been occupied than they 
otherwise would be. This is because they 
have a larger overlap with the excited state than the single  state with a positive 
$x$-component. This breaks the symmetry of reflection around the $z$-axis. 
This is evident in  
\frf{daryPRE}, where we can see the clear increase in the probability of 
the true state being in the PRE state on the far left and a substantial 
decrease in the probability of being the PRE state closest to the ground 
state. 

It should also be apparent, since the smoothed quantum state will be a 
mixture of the states in the PRE over the interval of interest, that the 
Bloch vector will lie within the triangle formed by the PRE states (the 
grey shaded area in \frf{Simp_Bloch}). As a result, the smoothed quantum 
state will not pass through the maximally mixed state (the centre of the 
circle) as the observed jump approaches. This is not to say that the 
length of the Bloch vector does not decrease, just that it does not go 
to zero. This can bee seen in \frf{Fig-Hyp3}.

\section{Proof of Result 3: Comparing the expected cost functions}\label{sec-R3}

As we have just seen in the last two 
examples, the commutativity of the filtered state and retrofiltered effect 
imposes no apparent constraints on the dynamics of the optimal 
smoothed quantum state. However, the fact that the optimal estimates 
in these cases are different gives rise to a new question. Alice 
cannot compute her optimal smoothed quantum state without knowledge of 
the nature of Bob's measurement, as this determines 
the set of possible true states. 
But what is the `best' measurement unravelling for Bob to perform, 
from Alice's point of view? 
Since we already have a metric,  the expected 
cost function, whose minimization defines the optimal estimate 
($\rho\fil$ or $\rho\sm$), it seems natural to use that expectation 
value as a measure of how good Alice's estimate is.
For our particular 
cost function, this is equivalent to the measurement that results in the 
purest smoothed state. 

\begin{figure}[t!]
\includegraphics[scale=0.28]{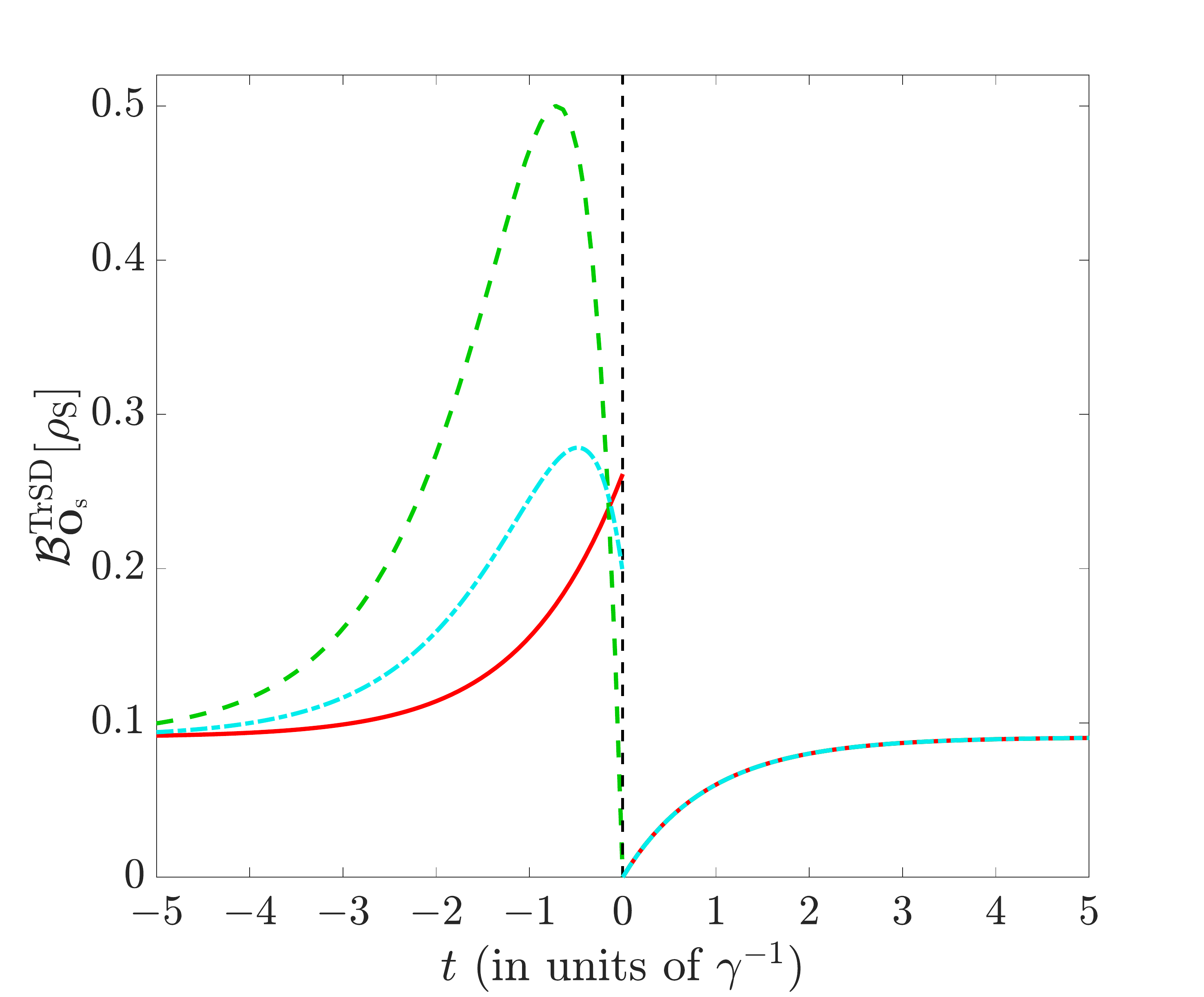}
\caption{The expected cost of the smoothed state when Alice's measurement 
record is obtained by photon detection and Bob's record is obtained by 
either photon detection (green dashed line), X-homodyne (red line) or by an
adaptive strategy based on weak local oscillators (cyan dot-dashed line) 
(see Sec.~\ref{ssec-adap}. The expected cost 
for both the X-homodyne case and the adaptive weak local oscillator cases 
have a lower expected cost than the photon detection (classical) case for 
most times due to the latter having to pass through the maximally mixed 
state to reach the (pure) excited state. See \frf{Fig-Hyp1} for the 
parameter details.}
\label{Fig-Hyp2}
\end{figure}

Now, intuitively, the measurement for Bob that would give the greatest 
purity would be the measurement that causes $\rho\sm(t) = \rho\sm\cl(t)$ 
as Alice only has to estimate between perfectly distinguishable true 
states. 
Applying this type of 
logic to the three cases that we have already considered, one would 
guess that after the case where Bob uses photon detection 
(having two perfectly distinguishable true states), the next best case would be 
 the adaptive scheme (with three non-orthogonal true states), followed by 
 the X-homodyne case (with a continuous infinity  
of non-orthogonal state to distinguish). Comparing the expected 
cost functions for each estimate in Fig.~\ref{Fig-Hyp2}, we can see that 
the above intuition is incorrect. In fact, the complete opposite holds 
over the majority of the time. This is despite the fact that 
the classically smoothed quantum state $\rho\sm\cl(t)$ is pure immediately 
prior to the jump occurring. Actually we have already discussed how this 
is linked with a prior decrease in purity in the classical case, in Sec.~\ref{ssec-photo}. 
With only two states, with the steady state being relatively close to the ground state,  
and with the state just prior to the jump being the excited state, 
the classical smoothed state must pass through the maximally mixed 
state, at which point it must be the worst (highest expected cost) estimator. 
This leads us to our final result, that 
the classically smoothed quantum state does not necessarily yield the 
lowest expected cost function.

We can gain a more general intuition as to why increasing the number of possible true 
states yields purer smoothed estimates, by analyzing the expression for 
the purity. That is,
\beq
\begin{split}
P\sm(t) &= \Tr[\rho\sm(t)^2] \\
&= \Tr\left[\left(\sum_{\rho\god\in{\mathbb T}_t} \wp(\rho\god;t|
\both{\bf O})\rho\god\right)^2\right]\\
&= \sum_{\rho\god,\rho\god'\in {\mathbb T}_t} \wp(\rho\god;t|\both{\bf O}) 
\wp(\rho\god';t|\both{\bf O}) \Tr[\rho\god \rho\god']\,,
\end{split}
\eeq
where for ease of illustration we have assumed a discrete set of possible 
true states.
We see that the purity is a weighted sum of the overlap 
between possible true states. Thus, 
with only two 
orthogonal true states in ${\mathbb T_t}$, the only terms that contribute 
are when $\rho\god = \rho\god'$. However, as the number of possible true 
states increase, 
additional terms that result from non-orthogonal states will 
also contribute, increasing the sum. Following this intuition, we arrive 
at the ordering that is observed in Fig.~\ref{Fig-Hyp2}, over 
the great bulk of times. 
The exception is when the jump is imminent, 
where the ordering flips. (From Fig.~\ref{Fig-Hyp2}, 
this flipping might appear to happen at an instant, but zooming in one finds 
that the three line do not intersect at the same point in time.) 
For the photon detection case the reason for the reversal is clear; as discussed 
above, just before Alice's jump the state is pure and so the expected cost is zero. 
Something similar, but less dramatic, happens with the adaptive jump case. 
Just before Alice's jump, one of the three possible true states becomes much more likely than the others, 
as discussed in Sec.~\ref{ssec-adap}. We refer the reader to Fig.~\ref{daryPRE} again to appreciate the difference 
from the homodyne case.

\blk As an aside, in Ref.~\cite{LGW-PRA21} it was shown that the trace-square 
deviation from the true state is not the only cost function that has the 
smoothed state as its optimal estimator, so does the relative entropy. 
One might then ask, does the classically smoothed state give the lowest 
expected relative entropy? The simple answer is no, in fact the ordering 
will not change. This is because, as shown in Ref.~\cite{LGW-PRA21}, when the true 
state is pure, the expected relative entropy between the smoothed state 
and the true state reduces to the von-Neumann entropy. Since, for qubits,
the von-Neumann entropy is a monotonic function of the purity, the ordering will remain.\blk

\section{Conclusion}\label{sec-concl}
In this paper we have proved three results regarding quantum state 
smoothing. These three results were motivated by the application of 
classical smoothing techniques in quantum experiments. These experiments 
assumed the unobserved information in the environment was 
made classical in 
a way that removed the `quantumness' 
from the system. Specifically, the quantum system was effectively replaced by 
a classical system, always in one of a set of orthogonal states with 
dynamics described by an hidden Markov model. 

One might say that these assumptions were physically reasonable in the context of these 
experiments. However, we argue that it is important to state an assumption like this 
explicitly, because in its absence one cannot justify the application of 
classical smoothing techniques for a quantum state. Our argument is \blk established by our first result, and is backed up \blk by two even stronger negative results, regarding the 
applicability or usefulness of classical smoothing techniques.
%
All three of the results in our paper were proven using a simple qubit system (atom) coupled 
to three decoherence channels. The measurements by the observer (Alice) has a 
classical description, describable in terms of transitions between the ground and excited states, and 
the master equation has no terms that generate coherence between these states. 

 Our first result was to show that the optimal smoothed estimate for this system in this situation depends 
 on how the information unobserved by Alice in the environment is assumed to become classical. 
In particular, we here 
considered two potential measurement unravellings for the environment, 
which for convenience we call Bob's measurement schemes. For 
photon detection, which is equivalent to assuming an orthogonal hidden Markov 
model for this system, the classical smoothed state (probability distribution over the two orthogonal states)
is recovered. But for homodyne detection a different smoothed state is found, albeit  
still a mixture of the states in the classical model. 

Our second result was that one 
cannot even make a statement on the diagonal basis of the smoothed quantum 
state without knowing the unravelling of the unobserved information. 
For this we considered a third type of measurement for Bob: an adaptive 
strategy using local oscillators and photon counters. 
Under this scheme, despite the apparent classicality of Alice's measurement, 
the smoothed quantum state is not, in general, a mixture of ground and excited states. 

Our third and final result is the most subtle. In the face of the first 
two results, one might still hope that the classical smoothing technique would 
be better than the other techniques; that the simplicity of the photon detection scheme for Bob, relative 
the others (homodyne and adaptive), would be reflected in Alice's ability to estimate the 
resulting unknown record and hence estimate the associated true state. (This ability is quantified 
by the trace squared deviation between the true state and the state estimate, which is the 
cost function whose minimization defines the optimal estimate.) Contrary to this intuition, 
we showed in our example that, most of the time, the expected cost is {\em higher} for the 
classical smoothing technique than for the quantum smoothing technique 
appropriate for the other unravellings.

Our results raise several questions for future research. 
\blk Whilst \blk we have shown that a classical description of Alice's filtering and retrofiltering is 
not sufficient for classical smoothing to be applicable in general, and that a classical 
description of the true state {\em is} sufficient, we do not know whether the latter 
condition is {\em necessary}, or whether some weaker condition would be sufficient. 
Another question is whether,  given a classical description of Alice's filtering and retrofiltering, 
there could be a different cost function such that the 
classically smoothed quantum state in \erf{eq-clqtsmooth} 
would be the optimal estimator for quantum systems.  
Another way that one might try to apply classical smoothing would be 
to diagonalize the filtered state at each instant in time and then compute 
new weightings for  those eigenstates using classical smoothing. 
Lastly, as mentioned in Ref.~\cite{CGW19}, it would be interesting to investigate what 
would happen if Alice assumed that Bob's unravelling had a classical description 
when in fact it does not. 
How poorly does this `wrong-guessed' smoothed state perform in terms of the 
trace-square distance, and could we find unravellings for 
Bob (or even Alice) that minimizes the deviation from the optimal expected 
cost?

\begin{acknowledgements}
This research is 
funded by the Australian Research Council Centre of Excellence Program, grant number  
CE170100012. K.T.L. was supported by an Australian 
Government Research Training Program (RTP) Scholarship. A.C.~acknowledges the support of the NSRF of Thailand via the Program Management Unit for Human Resources \& Institutional Development, Research and Innovation, grant number B05F650024
\end{acknowledgements}

\appendix
\begin{widetext}
\section{Co-diagonality of the filtered state and retrofiltered 
effect}\label{app-qest}
In this appendix we prove that when 
$\rho\god(t) \in {\mathbb T}_{\rm diag} = \{\ket{\psi_m}\bra{\psi_m}:
\langle{\psi_m}|\psi_{m'}\rangle = \delta_{m,m'} \}$, both the 
filtered state (provided the state is initially diagonal in this basis) and 
retrofiltered effect are necessarily diagonal in any 
basis that contains these states. However, before we present the proof, we 
need to express how both the true and filtered quantum states evolve in 
terms of quantum maps. For the true quantum state, a measurement operator \cite{BrePet06,WisMil10}
is assigned for each measurement outcome, which we will denote as 
$\hat{M}\ob(\by\ob;t)$, where $\by\ob(t)$ denotes the observed measurement 
outcome, and similarly for the unobserved measurement, be it a detector 
click or a photocurrent. With these measurement operators, the 
true state evolves as \cite{GueWis20}
\beq \label{True_Evo}
\tilde{\rho}\god(t+\dd t) = 
\hat{M}\ob(\by\ob;t)\hat{M}\un(\by\un;t)\tilde{\rho}\god(t)
\hat{M}\un\dg(\by\un;t)\hat{M}\ob\dg(\by\ob;t)\,.
\eeq
Here, for a later derivation, we have refrained from normalizing the state 
at each time step, as indicated by the tilde. Note, for notational simplicity we have absorbed 
the operator that generates the deterministic part of the evolution 
(including the Hamiltonian part of the evolution) into 
the observed measurement operator. 

For the filtered quantum state, we condition only on the observed 
measurement record, with the unobserved channel contributing the 
decoherence affecting the state. In general, one can recover the 
appropriate dynamics of the filtered state by averaging over the possible 
unobserved records. Averaging \erf{True_Evo} over the unobserved 
measurement record, the (unnormalized) filtered state evolves according to 
\beq\label{Filt_Evo}
\tilde{\rho}\fil(t+\dd t) = \sum_{\by\un(t)}\hat{M}\ob(\by\ob;t) 
\hat{M}\un(\by\un;t)\tilde{\rho}\fil(t)\hat{M}\un\dg(\by\un;t)
\hat{M}\ob\dg(\by\ob;t)\,,
\eeq
where we have assumed, for simplicity, that the spectrum of measurement 
outcomes is discrete. For a continuous spectrum this sum is replaced 
by an appropriate integral. To determine how the retrofiltered effect evolves, 
we use the fact that $\Tr[\tilde{\rho}\fil(t)\hat{E}\rfil(t)] = \tilde{\wp}
(\both{\bf O})$, which is independent of the time $t$. Thus, we have 
$\Tr[\tilde{\rho}\fil(t)\hat{E}\rfil(t)] = \Tr[\tilde{\rho}\fil
(t+\dd t)\hat{E}\rfil(t+\dd t)]$. Applying \erf{Filt_Evo}, the cyclic 
property of the trace and the fact that this holds for all possible 
filtered states gives us the dynamical equation for the retrofiltered 
effect, 
\beq\label{Retr_Evo}
\hat{E}\rfil(t) = \sum_{\by\un(t)}
\hat{M}\un\dg(\by\un;t)\hat{M}\ob\dg(\by\un;t)\hat{E}\rfil(t+\dd t)
\hat{M}\ob(\by\ob;t)\hat{M}\un(\by\un;t)\,.
\eeq

\underline{\bf Proof:} Given that $\rho\god(t) \in {\mathbb T}_{\rm diag}$ 
over the interval $[t_0,T)$, \erf{True_Evo} becomes
\beq\label{diag_True_Evo}
\ket{\psi_i}\bra{\psi_i} \propto 
\hat{M}\ob(\by\ob;t)\hat{M}\un(\by\un;t)\ket{\psi_j}\bra{\psi_j}
\hat{M}\un\dg(\by\un;t)\hat{M}\ob\dg(\by\ob;t)\,.
\eeq
By expressing the product of measurement operators as 
$\hat{M}\ob(\by\ob;t)\hat{M}\un(\by\un;t) = 
\sum_{i,j} T_{i,j}(\by\un;t) \ket{\psi_i}\bra{\psi_j}$, \erf{diag_True_Evo} 
places a restriction on the matrix elements $T_{i,j}(\by\un;t)$. Specifically, we 
have $T_{i,k}(\by\un;t)T^{*}_{k,j}(\by\un;t) \propto \delta_{i,j}$, where 
$A^*$ denotes the complex conjugate of $A$. Note, since we are 
only averaging over the unobserved measurement, we have dropped the 
dependence on $\by\ob(t)$ in the matrix elements for notational simplicity. 
Importantly, this constraint holds for all possible realizations of 
$\by\un(t)$ since the set ${\mathbb T}_{\rm diag}$ contains the possible 
true states for any realization of $\past{\bf O}_t$ and $\past{\bf U}_t$. 

With this constraint on the measurement operators, we can now compute the 
filtered state. Beginning with \erf{Filt_Evo}, we have
\beq
\tilde{\rho}\fil(t+\dd t) = \sum_{\by\un(t)}\sum_{i,j,k,\ell} T_{ij}
(\by\un;t)T^{*}_{k\ell}(\by\un;t) 
\ket{\psi_i}\bra{\psi_j}\tilde{\rho}\fil(t)\ket{\psi_k}\bra{\psi_\ell}\,.
\eeq
Now, assuming that the filtered state is diagonal at time $t$, \ie, 
$\tilde{\rho}\fil(t) = \sum_{m}\tilde{\wp}\fil(m;t)\ket{\psi_m}\bra{\psi_m}$, 
then 
\begin{align}
\tilde{\rho}\fil(t+\dd t) &= \sum_{\by\un(t)} \sum_{i,j,k,\ell,m}
T_{ij}(\by\un;t)T^{*}_{k\ell}(\by\un;t)\tilde{\wp}\fil(m;t) 
\ket{\psi_i}\langle\psi_j|\psi_m\rangle\langle\psi_m|\psi_k\rangle\bra{\psi_
\ell}\\
&= \sum_{\by\un(t)} \sum_{i,j,k,\ell,m}
T_{ij}(\by\un;t)T^{*}_{k\ell}(\by\un;t) \tilde{\wp}\fil(m;t) 
\delta_{j,m}\delta_{m,k}\ket{\psi_i}\bra{\psi_\ell}\\
&= \sum_{\by\un(t)} \sum_{i,j,\ell}T_{ij}(\by\un;t)T^{*}_{j\ell}
(\by\un;t)\tilde{\wp}\fil(j;t)\ket{\psi_i}\bra{\psi_\ell}\\
&= \sum_{\by\un(t)} \sum_{i,j,\ell} c_i(\by\un;t) \tilde{\wp}\fil(j;t) 
\delta_{i,\ell}\ket{\psi_i}\bra{\psi_\ell}\label{Sub_Filt}\\
&= \sum_{\by\un(t)} \sum_{i,j} c_i(\by\un;t)\tilde{\wp}\fil(j;t) 
\ket{\psi_i}\bra{\psi_i}\label{Sub_2_Filt}\,,\\
\end{align}
where, in the \erf{Sub_Filt}, we have substituted in the condition $T_{i,j}
(\by\un;t)T^{*}_{j,\ell}(\by\un;t) = c_i(\by\un;t) \delta_{i,\ell}$, 
where $c_i(\by\un;t)\in{\mathbb R}$ is the constant of proportionality.
Importantly, we see that the quantum maps that generate the evolution of 
the filtered state preserve the diagonality in the orthogonal basis 
$\{\ket{\psi_m}\}_m$. Thus, provided the initial condition is  
diagonal in the basis $\{\ket{\psi_m}\}_m$, the filtered state will remain diagonal 
while $\rho\god\in{\mathbb T}_{\rm diag}$. 

For the retrofiltered effect, the proof follows in a similar manner to that 
of the filtered state. Beginning by substituting the measurement operator 
into \erf{Retr_Evo} and assuming the retrofiltered effect at $t$ is 
diagonal in this basis, \ie, $\hat{E}\rfil(t+\dd t) = 
\sum_{m}E\rfil(m;t+\dd t)\ket{\psi_m}\bra{\psi_m}$, we have
\begin{align}
\hat{E}\rfil(t) &= \sum_{\by\un(t)} \sum_{i,j,k,\ell,m}
T^{*}_{ij}(\by\un;t)T_{k\ell}(\by\un;t)E\rfil(m;t + \dd t) 
\ket{\psi_i}\langle\psi_j|\psi_m\rangle\langle\psi_m|\psi_k\rangle\bra{\psi_
\ell}\\
&= \sum_{\by\un(t)} \sum_{i,j,\ell}T^{*}_{ij}(\by\un;t)T_{j\ell}
(\by\un;t)E\rfil(j;t+\dd t)\ket{\psi_i}\bra{\psi_\ell}\\
&= \sum_{\by\un(t)} \sum_{i,j,\ell} c_i(\by\un;t) E\rfil(j;t + \dd t) 
\delta_{i,\ell}\ket{\psi_i}\bra{\psi_\ell}\label{Sub_Retr}\\
&= \sum_{\by\un(t)} \sum_{i,j} c_i(\by\un;t)E\rfil(j;t + \dd t) 
\ket{\psi_i}\bra{\psi_i}\,,\\
\end{align}
where, in \erf{Sub_Retr}, we have used the fact that $T^{*}_{ij}
(\by\un;t)T_{j\ell}(\by\un;t) = (T_{ij}(\by\un;t)T^{*}_{j\ell}
(\by\un;t))^{*} = c_i(\by\un;t)\delta_{i,\ell}$. Once again, we see that 
the quantum maps that generate the evolution of the retrofiltered effect 
preserve the diagonality in the orthogonal basis $\{\ket{\psi_i}\}_i$. 
Thus, since $\hat{E}\rfil(T) \propto \hat{1}$, which is diagonal in every 
basis, the retrofiltered effect is necessarily diagonal when 
$\rho\god\in{\mathbb T}_{\rm diag}$.\\
\rightline{$\square$}

\end{widetext}

\section{Specification of the adaptive measurement scheme} 
\label{App:measScheme}

In this paper the physical system of interest is a qubit coupled to three decoherence channels: two emission channels (one monitored by Alice and one by Bob) and one absorption channel (monitored by Bob). In Sec.~\ref{ssec-adap}, the emission channel monitored by Alice is assumed to be very weak, $\delta\rightarrow 0^{+}$, meaning that, in the long time periods between her observed detections, her filtered state will reach a steady state.  In some parameter regimes~\cite{WarWis19} it is then possible for Bob to restrict the system state to a finite and time-independent set ${\mathbb T}$, which is known as a physically realizable ensemble (PRE). In order for this to occur, Bob must use a carefully chosen adaptive measurement scheme to monitor his emission and absorption channels. In Sec.~\ref{ssec-adap}, Bob is assumed to enact such a measurement scheme, which leads to the PRE of \frf{Simp_Bloch} when $\epsilon = 0.05\gamma$. \blk The PRE in question consists of three states ($\alpha$, $\beta$, $\gamma$) that are labelled by the angle of the Bloch vector in the $x$-$z$ great circle of the Bloch Sphere.
\blk

The freedom that Bob has when choosing his measurement scheme is most easily understood in a quantum optics context. Bob may utilise linear interferometers that take the field outputs of the system as inputs.  He may also use weak local oscillators (WLOs) that can be added to the interferometer outputs prior to photodetection. To achieve the PRE of Sec.~\ref{ssec-adap}, it turns out that Bob can employ a relatively simple measurement scheme that only uses WLOs and does {\em not} mix the system outputs.  The measurement scheme is further simplified as it obeys symmetries present in the unconditional evolution, namely that real-valued density matrices remain real-valued under the action of the ${\cal L}={\cal D}[\hat{\bf c}]$. Thus Bob separately adds a real-valued WLO to both his emission and absorption channels before photodetection.  The scheme is adaptive in the sense that the WLO amplitudes must be varied according to which state of the PRE is occupied.  

Before specifying the constraint equations that define the measurement scheme, we define the effective {\blk no-jump} Hamiltonian, $\hat{H}^{\prime}_{\rm eff}$, and jump operators, $\hat{\sigma}^{\prime}_{\mp}$, that are applicable when Bob is utilising real-valued WLOs. \blk The non-Hermitian operator $\hat{H}^{\prime}_{\rm eff}$ defines Bob's filtered system evolution in time increments when he observes no detections, whereas $\hat{\sigma}^{\prime}_{\mp}$ is proportional to the measurement operator used by Bob to evolve the state upon a detection from either of the two channels that he is monitoring. \blk 
\blk For example, if \blk the system state is currently $\ket{\alpha}$, then \blk
\begin{align}
\hat{H}^{\prime}_{\rm eff}(\alpha) &=-\frac{i}{2}\Bigl(\gamma\hat{\sigma}^{\dag}_{-}\hat{\sigma}_{-}
+\epsilon\hat{\sigma}^{\dag}_{+}\hat{\sigma}_{+}
+2\sqrt{\gamma}\alpha_{-}\hat{\sigma}_{-}   \nonumber \\
&  \hspace{6em} +2\sqrt{\epsilon}\alpha_{+}\hat{\sigma}_{+}
+\alpha_{-}^{2}+\alpha_{+}^{2}\Bigr)\\
\hat{\sigma}^{\prime}_{-}(\alpha)&= \hat{\sigma}_{-}+\alpha_{-}\\
\hat{\sigma}^{\prime}_{+}(\alpha)&=\hat{\sigma}_{+}+\alpha_{+}.
\end{align}
Here, $\alpha_{\mp}$ are the WLO amplitudes added to the emission and absorption channels, respectively. The jump and no-jump operators for the other states of the PRE are obtained simply through changing the WLO amplitudes 
to \blk $\phi_{\mp}$ and $\beta_{\mp}$. 
The values of all of these WLO amplitudes are \blk yet to be determined.

 The cyclic PRE is then defined by \blk requiring that each of the PRE states is an eigenstate of the appropriate no-jump operator:
\begin{align}
\hat{H}^{\prime}_{\rm eff}(\alpha)\ket{\alpha}&\propto \ket{\alpha}\label{cyclicHalpha}\\
\hat{H}^{\prime}_{\rm eff}(\phi)\ket{\phi}&\propto \ket{\phi}\label{cyclicHphi}\\
\hat{H}^{\prime}_{\rm eff}(\beta)\ket{\beta}&\propto \ket{\beta}\label{cyclicHbeta};
\end{align}
\blk and that the jump operators cyclically move \blk the state around the ensemble:
\begin{align}
\hat{\sigma}^{\prime}_{-}(\alpha)\ket{\alpha}&\propto 
\ket{\phi}, \ 
\hat{\sigma}^{\prime}_{+}(\alpha)\ket{\alpha}\propto\ket{\phi}
\label{cyclicJumpalphaP}\\
\hat{\sigma}^{\prime}_{-}(\phi)\ket{\phi}&\propto\ket{\beta}, \ 
\hat{\sigma}^{\prime}_{+}(\phi)\ket{\phi}\propto\ket{\beta}
\label{cyclicJumpphiP}\\
\hat{\sigma}^{\prime}_{-}(\beta)\ket{\beta}&\propto\ket{\alpha},\ 
\hat{\sigma}^{\prime}_{+}(\beta)\ket{\beta}\propto\ket{\alpha}
\label{cyclicJumpbetaP}
\end{align}

\blk Given that the states comprising the PRE are known, \erfs{cyclicHalpha}{cyclicJumpbetaP} can be easily solved \blk numerically for $\alpha_{\mp}$, $\phi_{\mp}$, and $\gamma_{\mp}$. \blk It is worth noting that rounding errors in the specification of the PRE states mean that a minimisation approach to solving the constraints is necessary (such as minimising the sum of the constraints squared).  Also important is that we have not specified how the PRE states are found in the first place, rather we have focused on the appropriate measurement scheme to achieve the PRE.  Techniques for finding the PRE are discussed elsewhere (See Ref.~\cite{WarWis19}), but here we \blk actually apply it, to derive the measurement scheme. \blk The values we find for $\alpha_{\mp}$, $\phi_{\mp}$, and $\gamma_{\mp}$ are as follows (with the upper and lower values matching the $-$ and $+$ subscripts): \blk 
\begin{align}
\alpha_{\mp}=
\begin{bmatrix}
-0.07812 \\
           -0.1804
         \end{bmatrix}, 
\ \phi_{\mp}=
\begin{bmatrix}
-0.3684 \\
           0.2552 
         \end{bmatrix}, 
\ \beta_{\mp}=
\begin{bmatrix}
           0.06446 \\
           0.6158
         \end{bmatrix}.
         \label{measScheme}
\end{align}
As an example, when Bob knows the system state is $\ket{\phi}$ he should add a WLO of amplitude $-0.3684$ to the emission channel and a WLO of amplitude $0.2552$ to the absorption channel. If he does this then the system will stay in the state $\ket{\phi}$ until he registers a detection. Upon registering a detection (from either of his two photodetectors) the system state will jump to $\ket{\beta}$.  Bob then should immediately switch the WLO amplitudes to $\beta_{\mp}$, which will stabilise the system in state $\ket{\beta}$.  In this way, Bob will know the system state and restrict it to the PRE ${\mathbb T}$. As a final point, more than one measurement scheme capable of realising the PRE of \frf{Simp_Bloch} was discovered in our investigation, with the scheme of \erf{measScheme} possessing the most symmetry.  This measurement scheme freedom may be explored further elsewhere.

\blk


\begin{thebibliography}{47}
\bibitem [{Com()}]{Comment1}%
  \BibitemOpen
  \href@noop {} {}\bibinfo {note} {While we consider the case where the
  environment unravels the remaining $1-\eta$ fraction, it is not necessary to
  the theory but rather stems from the context we are applying the smoothing
  theory to. In the scenario where there are two real observers, Alice and Bob,
  one can replace this fraction by Bob's measurement efficiency $\eta\un$,
  where Alice's and Bob's efficiency may not sum to unity.}\BibitemShut {Stop}%
\end{thebibliography}
\end{document}